\documentstyle[aps,prb,epsf]{revtex}
\newcommand{\be}{\begin{equation}}
\newcommand{\ee}{\end{equation}}
\newcommand{\ccre}[1]{c^{\dagger}_{{#1}}}

\newcommand{\can}[1]{c^{}_{{#1}}}

\newcommand{\cnum}[2]{c^{\dagger}_{{#1}}c^{}_{{#2}}}

\begin{document}
\title{Spectral Densities of Response Functions for  
the O(3) Symmetric  Anderson and Two Channel Kondo Models }
\author{S.C. Bradley$^1$,
R. Bulla$^2$, 
%
A.C. Hewson$^1$, and G-M. Zhang$^3$\\
%
%
$^1$Department of Mathematics, Imperial
 College, London SW7 2BZ, UK.\\
$^2$Theoretische Physik III, Elektronische Korrelationen
und Magnetismus, 
Universtit\"at Augsburg,
 86135 Augsburg, Germany\\
$^3$Centre for Advanced Study, Tsinghua University,
 Beijing 100084, P.R.
 of China}
%
%
\maketitle
\begin{abstract}
The O(3) symmetric  Anderson model is an
example of a system which has a stable low energy marginal Fermi liquid fixed
point for a certain choice of parameters. It is also exactly equivalent, in the large $U$ limit,
to a localized model which describes the spin degrees of freedom of the linear dispersion
two channel Kondo model. We first use an argument based on conformal field theory to establish 
this precise equivalence with the two channel model. We then use
the numerical renormalization
group (NRG) approach to calculate both one-electron and two-electron response functions
 for a range of values of the interaction strength 
$U$. We  compare the behaviours about the marginal Fermi liquid 
and Fermi liquid fixed points and interpret the results in terms of a renormalized Majorana
fermion picture of the elementary excitations. In the marginal Fermi liquid case the spectral
 densities of all
 the Majorana fermion modes display
a $|\omega|$ dependence on the lowest energy scale, and in addition
 the zero Majorana mode 
has a delta function contribution. The weight of this delta function is studied as a function
 of the interaction $U$ and is found to decrease exponentially with $U$ for large $U$.
Using the equivalence with the two channel Kondo model in the large $U$ limit,
we deduce the dynamical spin susceptibility of the two channel Kondo model over the full frequency range.
We use renormalized perturbation theory to interpret the results
and to calculate the coefficient of the
ln$\omega$ divergence found in the low frequency behaviour of the
$T=0$ dynamic susceptibility.
\end{abstract} 
\section{Introduction}
\label{intro}

The O(3) symmetric Anderson model is a modified form of the symmetric 
 Anderson model which was introduced by  
Coleman and Schofield  \cite{cs}. The Hamiltonian  can be expressed in the form,
$$ H=\sum_{n=0,\sigma}^{\infty}
t_n(\cnum{n+1,\sigma}{n\sigma}+\cnum{n\sigma}{n+1,\sigma})$$
\be +\sum_{\sigma}V_{\sigma}(
\cnum{0\sigma}{d\sigma}+\cnum{d\sigma}{0\sigma})+V_2(\ccre{d\downarrow}\ccre{0\downarrow}+
\can{0\downarrow}\can{d\downarrow})\ee
\be  +U(n_{d\uparrow}-{1\over 2})
(n_{d\downarrow}-{1\over 2}),\label{sam}\ee
where $\ccre{d\sigma}$ and $\can{d\sigma}$ are creation and annihilation operators 
for the localized impurity  d state with spin $\sigma$, with $V_{\uparrow}=V$,
$V_{\downarrow}=(V_0+V)/2$ and $V_{2}=(V_0-V)/2$.
 The electrons in the impurity state interact
via a Coulomb matrix element $U$, and are coupled by the hybridization terms to
the conduction electrons which are in
the form of a tight-binding chain with $c^{\dagger}_{n\sigma}$, $c^{}_{n\sigma}$,
the creation and annihilation operators at site $n$, and $t_n$ is the nearest neighbour
hopping matrix element. The model is basically a symmetric Anderson model ($V=V_0$) with an
anomalous hybridization term ($V\ne V_0$) which breaks both spin and charge conservation. It takes 
a simpler form when expressed in terms of Majorana fermions operators. For the impurity
d electrons 
 the Majorana
fermion operators are defined by
\begin{equation}d_1={1\over {\sqrt 2}}(c^{}_{d\uparrow}
+c^{\dagger}_{d\uparrow})\quad d_2={i\over {\sqrt 2}}(c^{}_{d\uparrow}
-c^{\dagger}_{d\uparrow}),\label{maj1}\end{equation}
\begin{equation} d_3={-1\over {\sqrt 2}}(c^{}_{d\downarrow}
+c^{\dagger}_{d\downarrow})\quad  d_0={i\over {\sqrt 2}}
(c^{}_{d\downarrow}-c^{\dagger}_{d\downarrow}),\label{maj2}\end{equation}
which satisfy the commutation relations,
\begin{equation}\{d_{\alpha}, d_{\beta}\}=
\delta_{\alpha,\beta},\label{cr}\end{equation}
where $\{\}$ indicates an anticommutator.\par
The Majorana operators for the conduction electrons are similarly 
defined by
\begin{eqnarray}
\chi_1(n)={e^{in\pi/2}\over {\sqrt 2}}((-1)^nc^{}_{n\uparrow}
+c^{\dagger}_{n\uparrow}),\nonumber \\ 
\chi_2(n)={ie^{in\pi/2}\over {\sqrt 2}}((-1)^nc^{}_{n\uparrow}
-c^{\dagger}_{n\uparrow}),\nonumber \\ 
\chi_3(n)={-e^{in\pi/2}\over {\sqrt 2}}((-1)^nc^{}_{n\downarrow}
+c^{\dagger}_{n\downarrow})\nonumber \\ 
\chi_0(n)={ie^{in\pi/2}\over {\sqrt 2}}
((-1)^nc^{}_{n\downarrow}-c^{\dagger}_{n\downarrow}),
\end{eqnarray}
with commutation relations as in (\ref{cr}).\par
 Using the Majorana fermion representation 
 the O(3) Anderson model can be expressed in the form      
$$H=i\sum_{\alpha=0}^3 V\chi_{\alpha}(0)d_{\alpha}
+iV_0\chi_{0}(0)d_{0}$$
\begin{equation}+i\sum_{\alpha=0}^3 \sum_{n=0}^{\infty}t_n\chi_{\alpha}(n+1)
\chi_{\alpha}(n)+Ud_{1}d_{2}d_{3}d_{0},\label{hamm}\end{equation}
which is clearly invariant under the O(3) group of transformations of the basis of the 
{1,2,3} Majorana fermions.\par  
There are two aspects of this
model that make it worthy of study.  One is that the model displays non-Fermi liquid behaviour
corresponding to marginal Fermi liquid theory in lowest order perturbation theory 
for $V_0=0 $ \cite{zh}. The marginal behaviour results from
singular scattering with an uncoupled local Majorana fermion mode. Numerical renormalization group
 (NRG) calculations \cite{bh} have established that
 this marginal Fermi liquid behaviour persists down
to $T=0$ and that there is an associated  residual entropy of ${\rm ln}2/2$.
 This fixed point  behaviour has been  interpreted in terms of the
free renormalized Majorana fermion modes.  The many-body excitations constructed from these
 free Majorana fermions account for the excitations found at the
 fixed point in the NRG calculations \cite{bhz}. 
 For $V_0\ne 0$ there is no uncoupled local
Majorana fermion mode and the
thermodynamic behaviour corresponds to Fermi liquid theory in terms of renormalized Majorana
fermions. \par

The original reason  Coleman and Schofield \cite{cs}  studied this model is because,
in  the large $U$ regime under a Schrieffer-Wolff transformation, it maps into a localized spin
model, the $\sigma$-$\tau$ model  \cite{cit}, which 
has a similar form to the two channel Kondo model (TCKM). The difference is that 
 the localized spin $S=1/2$ is coupled 
via an exchange interaction 
to the  spin and the 
isospin of conduction electrons in a single  channel rather than the spins of the conduction electrons
in two distinct channels as in the TCKM (for a comprehensive review of this model see \cite{cz}).
 The Hamiltonian of the model can be 
written in the form,
\begin{equation}H=\vec{S}_d \cdot[J_1\vec{\sigma}(0)+J_2\vec{\tau}(0)]+
\sum_{n=0,\sigma}^{\infty}t_n(\cnum{n+1,\sigma}{n\sigma}+\cnum{n,\sigma}{n+1,\sigma}),
\label{tsm}\ee
where $\vec{S}_d$ is the operator for the localized spin and  $J_1$, $J_2$ 
are the two exchange
 couplings.\par

The spin operators,
\begin{eqnarray} 
\sigma^{+}(n)=c^{\dagger}_{n\uparrow}c^{}_{n\downarrow},\quad \sigma^{-}(n)
=c^{\dagger}_{n\downarrow}c^{}_{n\uparrow}, \nonumber \\
 \sigma_{\rm z}(n)={1\over 2}
(c^{\dagger}_{n\uparrow}c^{}_{n\uparrow}-c^{\dagger}_{n\downarrow}c^{}_{n\downarrow}),
\label{spinop}
\end{eqnarray}
 in the basis spanned by the two singly occupied fermion states, 
 $|0(\uparrow), 1(\downarrow)\rangle$ and
$|1(\uparrow), 0(\downarrow)\rangle$, give a representation for the SU(2)
 algebra for spin $\sigma={1\over 2}$. The
isospin operators,
$$\tau^{+}(n)=(-1)^nc^{\dagger}_{n\uparrow}c^{\dagger}_{n\downarrow},
\quad \tau^{-}(n)=(-1)^nc^{}_{n\downarrow}c^{}_{n\uparrow},$$
\be \tau_{\rm z}(n)={1\over 2}
(c^{\dagger}_{n\uparrow}c^{}_{n\uparrow}+c^{\dagger}_{n\downarrow}
c^{}_{n\downarrow}-1),\label{isospinop}\ee
 give a  representation of the same algebra in the space spanned
 by the zero and double occupation states, $| 0,0\rangle$ and $| 1,1\rangle$.
By applying the Schrieffer-Wolff transformation  it can be shown
that  the $\sigma$-$\tau$ model corresponds to the localized
or large $U$ limit of the O(3) Anderson model 
with the coupling $J_1$ and $J_2$ given by
\be J_1={{2V(V+V_{0})}\over{U}}\quad J_2={{2V(V-V_{0})}
\over{U}}.\ee
The channel isotropic model $J_1=J_2=J$,
corresponds to  $V_0=0$ in the O(3) Anderson model. In this case 
one of the Majorana fermions modes is uncoupled and the  singular scattering of the
conduction electrons gives rise to
a marginal Fermi liquid fixed point. In terms of Majorana fermions the 
$\sigma$-$\tau$ model (\ref{sigtau}) takes
the form,
\begin{equation}H=i\sum_{\alpha=0}^3 \sum_{n=0}^{\infty}t_n\chi_{\alpha}(n+1)
\chi_{\alpha}(n)-{iJ\over 2}\vec{S}_d \cdot\vec{\chi}(0)\times\vec{\chi}(0),\label{sigtau}\end{equation}
where $\vec{\chi}(n)=(\chi_1(n),\chi_2(n), \chi_3(n))$.\par
 Coleman et al \cite{cit} put forward arguments 
that this one band localized model has the same fixed point and low energy behaviour
as the two channel Kondo model. A bosonization approach by Schofield \cite{sco} supported this 
conjecture, and so did a comparision of our numerical renormalization group results \cite{bh}
for the large $U$ O(3) Anderson model with the results for the two channel model.\par 
In the first  section of this paper we show that if we assume linear dispersion for the
conduction electrons the relationship 
between the  $\sigma$-$\tau$ model  and the spin degrees of freedom of the two
channel model becomes an {\it exact} one, and what is more  it is not confined purely
to the asymptotic low energy regime but applies to all relevant energy scales. In the
subsequent  sections we extend our earlier numerical renormalization group calculations \cite{bh}
for the O(3) Anderson model to the calculation the dynamical response functions at $T=0$ both
for the marginal Fermi liquid and Fermi liquid cases, and interpret these results using
 renormalized perturbation theory  for the Majorana fermions \cite{ach,bhz}.
Finally we use the equivalence of the O(3) Anderson model and the two channel Kondo
model in the large $U$ limit to deduce the form of the dynamic spin susceptibility
for the two channel Kondo model over all relevant energy scales. We use
 renormalized perturbation theory  to deduce
the coefficient of the ln$\omega$ divergence found in the low frequency behaviour of the
real part of the dynamic susceptibility at zero temperature.\par

\section{Exact Equivalence to the Two Channel Kondo Model }

 We begin by showing  that the 
Majorana fermion description, used to formulate the O(3) Anderson model
and  $\sigma$-$\tau$ models,  can emerge quite naturally for the two channel 
model as a representation of the algebra of the total spin current of the two 
channels. For linear dispersion and no cut-off
we can express the Hamiltonian of the channel isotropic  TCKM model  
 in the form,
\begin{eqnarray}
 && H= H_0+H_I  \nonumber \\ 
 && H_0=\frac{v_f}{2\pi}\sum_{j=1}^{2}\sum_{\sigma=\uparrow,\downarrow}
    \int_{-\infty}^{+\infty}dx 
    :\psi^{\dag}_{j,\sigma}(x)(i\partial_x)\psi_{j,\sigma}(x): 
\nonumber \\ 
&& H_I=\sum_{a=x,y,z}J_a S_d^a J_s^a(0),
\end{eqnarray}
on taking a Fourier transform to a continuum variable $x$, where $v_f$ is the
Fermi velocity, $j$ is a channel index, and the conduction electron operators between the colons
 have to be to normal ordered.
We have assumed s-wave scattering only and replaced the incoming and
 outgoing waves with two 
left-moving electron fields $\psi_{j,\sigma}(x)$; $J_s^a(x)$ are the 
conduction electron spin current operators
\begin{equation}
   J_s^a(x)= \sum_{j,\sigma,\sigma'}
     :\psi^{\dag}_{j,\sigma}(x)s^a_{\sigma,\sigma'}\psi_{j,\sigma'}(x):  
\end{equation}
$s^a$ being spin-1/2 matrices. We can also introduce charge and flavour
currents
\begin{eqnarray}
&& J_c(x)=\sum_{j,\sigma}:\psi^{\dag}_{j,\sigma}(x)\psi_{j,\sigma}(x): 
\nonumber \\ 
&& J_f^a(x)=\sum_{j,j',\sigma}:\psi^{\dag}_{j,\sigma}(x)t_{j,j'}^a 
                     \psi_{j',\sigma}(x):.  
\end{eqnarray}
where $t_{j,j'}^a$ are generators of an SU(2) symmetry group. Following Affleck
and Ludwig \cite{al}, the free part of the Hamiltonian can be rewritten as a 
sum of three commuting terms by the usual point-splitting procedure
(Sugawara construction): 
\begin{eqnarray}\label{suga}
 H_0=\frac{v_f}{2\pi}\int_{-\infty}^{+\infty}dx
    && \left [\frac{1}{8}:J_c(x)J_c(x):+
     \frac{1}{4}:\vec{J}_f(x)\cdot\vec{J}_f(x):\right.
\nonumber \\       
     && \left .+\frac{1}{4}:\vec{J}_s(x)\cdot\vec{J}_s(x): \right ],
\end{eqnarray}
while the interaction term is expressed in terms of the electron spin 
currents and the impurity spin only. The information about the number of 
channels is contained in the commutation relations obeyed by the spin currents
\begin{equation}
 [J_s^a(x), J_s^b(x')]
  =i\epsilon^{abc}J_s^a(x)\delta(x-x')+\frac{ki}{4\pi}\delta_{a,b}\delta'(x-x')
\end{equation}
indicating that $J_s^a(x)$ form an SU(2) level $k=2$ Kac-Moody algebra. 
Meanwhile, the charge and flavour currents satisfy 
\begin{eqnarray}
&& [J_c(x),J_c(x')]=2ki\delta'(x-x'), \nonumber \\
&&  [J_f^a(x), J_f^b(x')]
  = i\epsilon^{abc}J_f^a(x)\delta(x-x')
 \nonumber \\
  && \hspace{3cm} +\frac{ki}{4\pi}\delta_{a,b}\delta'(x-x').
\end{eqnarray}
They form a U(1) Kac-Moody and another SU(2) level-2 Kac-Moody algebra, 
separately.

It is now quite natural to introduce a Majorana representation of the
spin current operators in the form,
\begin{eqnarray}
 && J_s^x(x)=-i:\chi_2(x)\chi_3(x):, \nonumber \\
 && J_s^y(x)=-i:\chi_3(x)\chi_1(x):, \nonumber \\
 && J_s^z(x)=-i:\chi_1(x)\chi_2(x):,
\end{eqnarray}
where $\chi_1(x),\chi_2(x),$ and $\chi_3(x)$ are left-moving free Majorana
fermion fields. It can be shown that this representation reproduces the SU(2) level-2 Kac-Moody
commutation relations. Our approach differs from earlier conformal field theory approaches 
to the TCKM \cite{ml,ye,gs}, where Majorana fermions are introduced only at a later stage after bosonization.
It is important to note that this Majorana representation of the spin currents 
is only appropriate
for the two channel model as it leads to a level-2 algebra. It would be
{\it inappropriate} for the single channel Kondo model where the corresponding
spin current generates a level-1 algebra.

In a similar way, we can also introduce Majorana representations for the
flavour currents
\begin{eqnarray}
 && J_f^x(x)=-i:\chi'_2(x)\chi'_3(x):, \nonumber \\
 && J_f^y(x)=-i:\chi'_3(x)\chi'_1(x):, \nonumber \\
 && J_f^z(x)=-i:\chi'_1(x)\chi'_2(x):,
\end{eqnarray}
which reproduces the commutation relations satisfied by the flavour currents,
and
\begin{equation}
  J_c(x)=-2i:\chi'_4(x)\chi'_5(x):
\end{equation}
can represent the charge current operator. Note that $\chi'_{\alpha}$ with
$\alpha=1,2,3,4,5$ are also left-moving free Majorana fermion fields.
It is well-known that the dynamics of charge, flavour, and spin are 
completely determined by the commutation relations of 
the current operators. Though the spin currents of the two channel Kondo model
can be represented in terms of three Majorana fermion fields
$\chi_{\alpha}(x)$ ($\alpha=1,2,3$), we emphasize that
they can not be given any simple physical interpretation in terms of the 
original conduction electrons $\psi_{j,\sigma}(x)$.

At this point we have the current operator terms in the Hamiltonian
as quartic in the Majorana fields. The Sugawara construction enables one to
write kinetic energy terms, which are quadratic in field operators, as quartic
terms. This is what was done earlier in writing the free part of the
Hamiltonian in form of equation (\ref{suga}), and it is convenient if one is 
pursuing a purely
algebraic approach as used in the conformal field theory \cite{al}. However
for our purposes it is more convenient now to perform an inverse Sugawara
construction by the usual point-splitting procedure again, and rewrite the
terms quartic in the Majorana fermions as  kinetic energy terms which are
quadratic \cite{kz,ginsparg}:
\begin{eqnarray}
&& :J_c(x)J_c(x):
   =4\sum_{\alpha=4}^{5}:\chi'_{\alpha}(i\partial_x)\chi'_{\alpha}(x);
\nonumber \\
&& :\vec{J}_f(x)\cdot\vec{J}_f(x):
   =2\sum_{\alpha=1}^{3}:\chi'_{\alpha}(i\partial_x)\chi'_{\alpha}(x);
\nonumber \\
&& :\vec{J}_s(x)\cdot\vec{J}_s(x):
   =2\sum_{\alpha=1}^{3}:\chi_{\alpha}(i\partial_x)\chi_{\alpha}(x).
\end{eqnarray}
The model Hamiltonian is transformed and divided into the following two parts,
\begin{eqnarray}
&& H_c+H_f=\frac{v_f}{4\pi}\sum_{\alpha=1}^{5}
  \int_{-\infty}^{+\infty}dx :\chi'_{\alpha}(x)(i\partial_x)\chi'_{\alpha}(x):,
\nonumber \\
&& H_s=\frac{v_f}{4\pi}\sum_{\alpha=1}^{3}
    \int_{-\infty}^{+\infty}dx :\chi_{\alpha}(x)(i\partial_x)\chi_{\alpha}(x):
\nonumber \\ && \hspace{3cm}        
      -\frac{iJ}{2}\vec{S}_d \cdot :\vec{\chi}(0)\times\vec{\chi}(0):.\label{hs}
\end{eqnarray}
$H_c+H_f$ describes the non-interacting charge and flavour degrees of freedom.
It has a symmetry group $ U(1)\otimes SU(2)_2 = SO(5) $ and is expressed by
five free Majorana fermion fields $\chi'_{\alpha}(x)$
($\alpha=1,2,3,4,5$). $H_s$ is the main part of the model and describes
the spin degrees of freedom with three left-moving Majorana 
fermion fields $\chi_{\alpha}$ ($\alpha=1,2,3$) interacting with the impurity 
spin. It has the symmetry $SU(2)_2$ or $SO(3)$ 
so that the full Hamiltonian has the symmetry group
$SO(5)\otimes SO(3)$, which is represented by eight different Majorana 
fermion fields.  

In the two channel model Hamiltonian, $H_s$ given in equation (24)  is the only part 
which includes an interaction with the impurity spin. This part of the 
Hamiltonian is exactly 
equivalent to the vector part of the $\sigma$-$\tau$ model defined
in equation (\ref{sigtau}).
The zero Majorana fermion contribution to (\ref{sigtau}) is completely
decoupled from the other modes so we can separate it out, and  
if we take the continuum limit with linear dispersion equation (\ref{sigtau})
can be written in the form,
\begin{eqnarray}
&& H=H_{\rm sc}+H_{\rm vec} \nonumber \\
&& H_{\rm sc}=\frac{v_f}{2\pi}
  \int_{-\infty}^{+\infty}dx :\chi_0(x)(i\partial_x)\chi_0(x):
\nonumber \\
&& H_{\rm vec}=\frac{v_f}{2\pi}\sum_{\alpha=1}^{3}
    \int_{-\infty}^{+\infty}dx :\chi_{\alpha}(x)(i\partial_x)\chi_{\alpha}(x):
\nonumber \\ && \hspace{3cm}
      -{iJ\over 2}\vec{S}_d \cdot :\vec{\chi}(0)\times\vec{\chi}(0):.
\end{eqnarray}
and can identify $H_{\rm vec}$ with $H_s$. This implies that the application of 
the $\sigma$-$\tau$ model is not restricted to the very low energy regime but 
 can be used to calculate the impurity contribution to the thermodynamics
of the two channel Kondo model {\it over the full temperature range}. This result is
{\it exact} subject only to the requirement of linear dispersion for the 
conduction electrons (this is not evident in the approaches that use bosonization
\cite{sco,ml,ye,gs}). The result implies that the O(3) Anderson model
in the large $U$ regime can be used to calculate the spin correlation functions
of the two channel Kondo model over the complete temperature and frequency range,
 $\omega, T << D$, where $D$ is the cut-off imposed on the conduction electrons.
We exploit this equivalence to calculate the dynamic susceptibility of the TCKM
in the last section of the paper.
Though we have restricted our derivation to the channel isotropic case $J_1=J_2=J$,
the result can be generalized to the channel anisotropic case $J_1\ne J_2$.\par

The fact that we have been able to 
establish the exact equivalence  the linear dispersion 
TCKM and  $\sigma$-$\tau$ 
models, not just for the low energy regime but over the full parameter
range of the model, is at first sight a little surprising. 
A feature of the general TCKM is that there is a 
strong coupling regime in which the impurity can be overscreened,
and this strong coupling fixed point was shown by Nozi\'eres and Blandin
\cite{nb} to be an unstable one. Impurity overscreening on the other hand
cannot occur in the strong coupling regime of the $\sigma$-$\tau$
model as the spin and isospin are mutually exclusive channels 
and  they cannot both screen the impurity simultaneously. The
strong coupling limit for the $\sigma$-$\tau$ model, in apparent
contrast to that of the two channel model, gives
  a stable fixed point.
However, to access the overscreened states  of the standard TCKM requires the coupling
$J$ to be much greater than the bandwidth $D$. In the linear dispersion
two channel model $D\to\infty$ so these states are not accessible in this
version of the two channel model. We must distinguish two stong coupling limits.
One in which the limit $J\to \infty$ is taken with a finite cut-off or before the limit $D\to\infty$
 is taken, and one in which the limits
are taken in the opposite order. The first leads to overscreening and
an unstable fixed point whereas the latter leads to a screened stable
fixed point.\par
A parallel situation arises in the standard Anderson model. 
If one assumes linear dispersion then the parameter regime in which the impurity
level $\epsilon_d$ lies outside the conduction band cannot be accessed.
The model with the level within the conduction band, and the one
with the level below the conduction band, in the Anderson model case
have the same fixed point but they do have different expressions
for the Kondo temperature (see \cite{achb}). \par
Before leaving this topic 
we briefly show that, because the Majorana representation of the TCKM  
is equivalent to the large $U$ limit of the O(3) Anderson model, we can use
results for the O(3) model to 
confirm the description of the TCKM fixed point as
one corresponding to renormalized asymptotically free 
Majorana fermions \cite{cit,ml,bhz,ye,gs}. Our earlier numerical renormalization
group calculations \cite{bh} established that the fixed point 
of the O(3) Anderson model corresponds to free Majorana
fermions independent of $U$.
This implies that the fixed point 
corresponds to the non-interacting model $U=0$.  
The many-body excitations and their degeneracies  
at the fixed point, as found in the NRG  
calculations, can be explained in terms of the many body excitations
built up from  the three non-interacting hybridized 1,2,3 Majorana fermion modes,
combined  with the excitations arising from the unhybridized zero Majorana
fermion modes described by $H_{\rm sc}$.
This interpretation of the fixed point in terms of the O(3) Anderson model is
simpler than the equivalent interpretation 
as the strong coupling limit
of the $\sigma$-$\tau$ model.  
In the latter approach a change
of boundary condition for the 1,2,3 Majorana fermion modes, 
which interact with the impurity,
has to be invoked as a result of the strong coupling, relative to that 
of the uncoupled 0 Majorana fermion mode.
This is not necessary in the Anderson model description
as there is naturally a difference of boundary
condition at impurity in the hybridized and unhybridized
terms. The leading (dangerously) irrelevent interaction term, which
leads to the logarithmically divergent contributions to the specific
heat coefficient and susceptibility in these models,
can be interpreted as the renormalized interaction term $\bar U$
in the Anderson model
(see \cite{bhz,cit} for further details). \par

To derive a complete description of the many-body excitations and their
degeneracies at the
fixed point for the TCKM from the $U=0$ O(3) Anderson 
model we simply have to replace  the uncoupled 0 Majorana degree of freedom
described by $H_{\rm sc}$ by the five uncoupled Majorana degrees of freedom
in $H_{\rm c}$ and $H_{\rm f}$,  which describe the charge and flavour 
excitations. There are two slightly different sets of excitations,
one set corresponding to a chain of even sites (sector A)
and another set corresponding to an odd number of sites (sector B).
When these are combined, as shown in Table 1, they reproduce the 
many-body energy levels and degeneracies found in the NRG calculations \cite{pc} 
and conformal field theory calculations \cite{al}  for the TCKM.
Since the interaction is restricted to the spin part of the Hamiltonian
$H_{\rm s}$ the leading (dangerously) irrelevent interaction which leads
to logarithmically divergent contributions to the specific
heat coefficient and susceptibility in the TCKM is exactly the same
 as the renormalized
interaction $\bar U\sim T_{\rm K}$ of the O(3) 
Anderson model, where $T_{\rm K}$ is the Kondo temperature.
We will use this explicitly later in section 4 to calculate  the dynamic susceptibility
of the TCKM from that of the O(3) Anderson model.

\section{Numerical Renormalization Group Calculations }\par
In this section we extend the numerical renormalization group method to calculate the 
dynamic features of the O(3) Anderson model.
For the basic details of the 
NRG method as applied to  this model we refer to the earlier paper \cite{bh} where the thermodynamic 
properties were calculated. Calculation of the dynamic behaviour via the NRG 
approach is rather more involved because additional matrix elements have also to be calculated in the
sequence of iterative diagonalizations.  The general details of the
NRG approach to the calculation of spectra for impurity models are given in 
 \cite{ssk,chz}. 
Because both charge and spin are not conserved for this model the calculations here differ from
the standard case in a number of respects and the relevant details are given in the appendix.\par
\bigskip
\subsection{Spectral densities of single particle Green's functions}\par
 \bigskip  
The spectral density $A_{\sigma}(\omega)$ of the single particle Green's function of spin $\sigma$
at  the 
impurity site
is given by
\begin{eqnarray}
   A_{\sigma}(\omega) = \frac{1}{Z}\sum_{nm}  
           \bigg\vert   \Big< n \Big\vert c^\dagger_{d\sigma}
                                         \Big\vert m \Big>
                   \bigg\vert^2
                   \delta \big( \omega -(E_{n} -E_{m}) \big) \nonumber \\
      \times     \left( e^{-\beta E_m} + e^{-\beta E_n} \right)
 \label{specd}
\end{eqnarray}
where $E_n$ are the exact eigenvalues of the Hamiltonian  and $Z$ is the partition function
 ($Z=\sum_n {\rm exp}{(-\beta E_n)}$). This is evaluated using the many-body states calculated
in the sequence of iterative diagonalizations of the NRG. Because of the decreasing
energy scales in the sequence of steps in the NRG, the spectral density for a particular
frequency  have to be calculated from the step in the sequence corresponding to the 
appropriate energy scale. The spectrum is then built up from the complete set of
NRG steps. Though the method has been developed for finite temperature situations, the procedure
is more straightforward and more accurate for the case $T\to 0$, and
we have restricted the calculations here to the low temperature limit.  Though a large number of
 states are retained at each
step in the NRG calculation (of the order of 300-500 not including degeneracies), the spectrum
is discrete and to obtain a continuous curve a broadening is introduced which decreases
logarithmically with the energy scale. The advantages of this approach over alternative methods,
such a Monte Carlo, are that it can be used at very low temperatures, can be used for any value of
$U$, and the results can be checked against known sum rules and identities. The details of
the low energy spectrum can be calculated to a very high resolution. For further details of the
approach we refer to the papers cited earlier \cite{ssk,chz}.\par
 Due to the anomalous hybridization term, the O(3) Anderson model  does not conserve either
total spin or total charge. However, the combination of total spin plus total isospin ${\bf T}$ is 
conserved and the many-body states and the eigenstates $|m\rangle$ can be classified using the
quantum numbers $T$ and $T_z$ associated with the total of this angular momentum  and its 
z-component (note that in reference \cite{bhz} the symbol ${\bf j}$ was used for this quantity,
but here we revert to the earlier notation ${\bf T}$ used in \cite{bh,bbh}). In evaluating
(\ref{specd}) the sum over the degeneracies associated with $T_z$ can be done explicitly 
as explained in the appendix. The single particle Green's functions for the impurity
can be expressed as linear combinations of the Green's functions for the Majorana fermions,
$G_{\alpha\beta}(t-t^{\prime})=-i\theta(t-t^{\prime})\langle 
\{d_{\alpha}(t)d_{\beta}(t^{\prime})\}\rangle$, for $\alpha,\beta=0,1,2,3$ using equations
(\ref{maj1}) and (\ref{maj2}). In the absence of
applied field these Green's
functions are diagonal ($\propto\delta_{\alpha,\beta}$), and due to the O(3) symmetry those corresponding
 to 
$\alpha=1,2,3$ are equal. We can deduce the corresponding spectral densities, 
$\rho_{\alpha}(\omega)$, using equation (\ref{specd}).  The results for these spectra for $V_0=0$, the
 non-Fermi-liquid case, are shown in figures 1-4 for a range of values
of the interaction $U$, and  corresponding results for $V_0\ne 0$ are shown in figures 8-10.\par
   In figure 1 the spectral density for the uncoupled $\alpha=0$  Majorana fermion (scalar) is shown
 together with those for the $\alpha=1,2,3$  Majorana fermions (vector) which are hybridized with
the conduction electrons (and which are all identical). There is an additional delta function
 contribution to the $\alpha=0$ spectral density which is not shown but its weight as a function
 of $U$ is shown in figure 6. What is clear is the cusp-like peak 
 in the spectral density of the scalar Majorana fermion at Fermi-level and the fact that the height
 of this peak is $U$ dependent. The spectral density for the vector Majorana fermions  also  has a cusp-like
peak at the Fermi-level but, in contrast to the scalar case, its value at the Fermi level is independent
 of $U$. In figure 2
the spectral densities of the scalar ($\alpha=0$) Majorana fermion are shown on their own with an inset
showing the linear behaviour in the neighbourhood of the Fermi level. The corresponding plots for
the vector Majorana fermions are given in figure 3, which also show linear behaviour in the 
neighbourhood of the Fermi-level.
 \par
We can analyze these results using the renormalized perturbation theory developed in earlier
work \cite{bhz,ach}. The perturbation theory is in terms of renormalized propagators
for the Majorana fermions, and renormalized parameters, $\bar \Delta=\bar z\Delta$, 
$\bar \Delta_0=\bar z_0\Delta_0$, and $\bar U=\sqrt{\bar z_0{\bar z}^3}\Gamma_{0123}(0,0,0,0)$,
where  $\bar z$, $\bar z_0$,are the wavefunction renormalization factors derived from the
non-singular contributions to the
self energies, $\Sigma(\omega)$ and $\Sigma_0(\omega)$, associated with the $\alpha=1,2,3$
and $\alpha=0$ Majorana fermions respectively, $\Delta_{\alpha}(\omega)=\pi
V_{\alpha}^2\rho_0(\omega)$ where $\rho_0(\omega)$ is the density of states of the
conduction electrons,
and 
 $\Gamma_{0123}(\omega_1,\omega_2,\omega_3,\omega_4)$,  is the irreducible four vertex
for the Majorana interaction. 
The retarded Green's function for the Majorana fermions then has the form,                                 
\be G_{\alpha\alpha}(\omega)={\bar z_{\alpha}\over {\omega+i\bar\Delta_{\alpha}-
\bar \Sigma^{(r)}_{\alpha\alpha}(\omega)-\bar \Sigma^{(s)}_{\alpha\alpha}(\omega)}},\label{rgf}\ee
where $\bar\Sigma^{(r)}_{\alpha}(\omega)$  and $\bar\Sigma^{(s)}_{\alpha}(\omega)$
are the regular and singular renormalized self-energies, and 
we have taken the wide band limit with $\Delta_{\alpha}(\omega)$ independent of
$\omega$. In the renormalized perturbation expansion there are counterterms
which have to be included to satisfy the renormalization conditions and the
Majorana fields are rescaled to absorb the wavefunction renormalization factors
so that the free propagators in the expansion are $1/(\omega+i\bar\Delta_{\alpha})$ 
(see \cite{bhz} for further details). \par
For the non-Fermi liquid case $V_0=0$ the  leading order low frequency  correction to the 
renormalized self-energy $\bar\Sigma(\omega)$
for the $\alpha=1,2,3$ Majorana fermions comes from the second order diagram for the
singular part $\bar\Sigma^{(s)}(\omega)$ shown in
figure 5(a). The result is
\be {\rm Re}\bar\Sigma^{(s)}(\omega)=\left({\bar U\over{\pi\bar\Delta}}\right)^2\omega{\rm ln}
\left({\omega\over{\bar\Delta}}\right),\ee
and
\be {\rm Im}\bar\Sigma^{(s)}(\omega)=
-{{\pi}\over 2}\left({\bar U\over{\pi\bar\Delta}}\right)^2{{|\omega|}\over{\bar\Delta}}\ee
for $T=0$.
As $U\to 0$, $\bar\Delta_{\alpha}\to \Delta_{\alpha}$, $\bar U\to U$ and this result
 goes over to that of the standard second order perturbation theory.
For the density of states of the $\alpha=1,2,3$ Majorana fermions in the
neighbourhood of the Fermi level we find,
\be \rho_{\alpha}(\omega)={1\over{\pi\Delta}}
\left[1-{{\pi}\over 2}\left({\bar U\over{\pi\bar\Delta}}\right)^2|\omega|+
{\rm O}(\omega^2{\rm ln}^2(\omega))\right],\label{rhovec}\ee
for $\alpha =1,2,3$. We see from this result that the density of states at the Fermi level is $1/\pi\Delta$,
independent of $U$, which is the same result as for the O(4) Anderson model,
and that the leading correction term is linear in $|\omega|$. These results 
explain the features seen in the plots of the spectral densities of the vector Majorana  shown in figures 1
 and 3. The linear term in the spectral density at the Fermi level is seen clearly and
the value at the Fermi-level agrees with the theoretical
value $1/\pi\Delta$ to within 2\%.\par
The second order contribution to the renormalized self energy $\bar\Sigma_0(\omega)$
for the $\alpha=0$ Majorana fermion is the same as the for the O(4) model as it only
involves explicitly the propagators of the renormalized $\alpha=1,2,3$ Majorana fermions (see figure 5(b)).
To lowest order the result for the imaginary part is
 \be {\rm Im}\bar\Sigma_0(\omega)=
-{{\bar \Delta}\over 2}\left({\bar U\over{\pi\bar\Delta}}\right)^2\left({\omega\over{\bar\Delta}}
\right)^2\ee
for $T=0$. The leading term for the real part of the self-energy is linear in $\omega$ but is cancelled by 
the counterterm contribution to comply with the renormalization conditions. When this is
inserted into equation (\ref{rgf}) then we find for the corresponding spectral density
\be \rho_{0}(\omega)=z_0\delta(\omega)+
{z_0\over {2\pi\bar\Delta}}\left({\bar U\over{\pi\bar\Delta}}\right)^2+
{\rm O}(\omega^2),\label{rho0}\ee
The delta function contribution at zero frequency reflects the fact that the zero mode 
associated with the uncoupled $\alpha=0$ local Majorana fermion exists down to $T=0$,
and it is this mode that leads to the anomalous entropy of ${\rm ln}(2)/2$. Its weight
in the spectral density is reduced to $z_0$, which we cannot calculate within the renormalized
perturbation approach. However, we can calculate it to leading order for small $U/\pi\Delta$,
and the result is $z_0=1/(1+(U/\pi\Delta)^2)$. For large $U/\pi\Delta$, the model can be related to
the two channel Kondo model and there is only one low energy scale, which is the Kondo temperature.
We expect, therefore, in this limit that $z_0\sim T_{\rm K}/\Delta$ which implies that for
large $U/\pi\Delta$, that $z_0$ should tend to zero exponentially in the variable $U/\pi\Delta$.
This behaviour is consistent with the NRG results for the delta function weight plotted
as a function of $U/\pi\Delta$ shown in figure 6. The non-delta function contribution 
has a finite limit as $\omega\to 0$ given in renormalized perturbation theory by the
the second term in equation (\ref{rho0}). 
 Unlike the spectra for the $\alpha=1,2,3$
Majorana fermions the value of this contribution at the Fermi level depends on $U$ and
monotonically increases with increase of $U/\pi\Delta$.  We see from equation (\ref{rho0})
that to lowest order in $U/\pi\Delta$ it is proportional to $U^2/(\pi\Delta)^3$.
As the second order diagram
in the renormalized perturbation theory gives the exact result to order $\omega^2$
for ${\rm Im}\bar\Sigma_0(\omega)$ there should be no further contributions 
to the leading two terms in equation (\ref{rho0}). In the large $U$ limit we
deduced from a comparison of the results with those for the two channel Kondo model
(see \cite{bhz}) that $\bar U/\pi\bar\Delta\to 1$. The height of the peak at the Fermi-level compared
with that for the vector term in this limit is then given by $\bar z_0/2\bar z$  (using $\bar\Delta
=\bar z\Delta$), and provides an estimate of the ratio of the wavefunction renormalization factors of the
scalar and vector Majorana fermions. \par
The correction terms in (\ref{rho0})
arise from terms in the imaginary part of the self-energy which behave as $\omega^4$
for small $\omega$. The second order diagram in the renormalized perturbation theory
gives an contribution $\omega^4$ but there are also contributions to this order from
higher order diagrams.  More important, however, are contributions to the imaginary part
which  behave like $\omega^2|\omega|$, as they result in corrections to (\ref{rho0})
which behave like $|\omega|$.  Such a term is generated from the higher order diagram
shown  in figure 7. The asymptotic contribution
to the spectral density from this diagram   as $\omega\to\ 0$ is 
\be -2z_0\pi^2\left({{\bar U}\over{\pi\bar\Delta}}\right)^4{ {|\omega|}\over {\bar\Delta^2}}.\ee
There may be other higher order diagrams which contribute terms to this order  $\omega$
but checking this is difficult. The explicit evaluation of the higher order diagrams is in general
complicated and there does not seem to be any simple way of seeing whether or not a diagram
gives a contribution of this order. It is clear from the results shown in figure 2  that there
is a net  contribution of this order as there is clear $|\omega|$  dependence in the neighbourhood 
of the Fermi level.  If $U$ is small compared to $\pi\Delta$ then the second order diagram
which gives an $\omega^2$ correction to (\ref{rho0})  should dominate over the leading $|\omega|$
term which is of fourth order. The second order result
corresponding to (\ref{rho0})  is  
\be \rho_{0}(\omega)=\left(1-\left({{ U}\over{\pi\Delta}}\right)^2\right)\delta(\omega)+
{1\over {2\pi\Delta}}\left({U\over
{\pi\Delta}}\right)^2\left(1-{{\omega^2}\over {2\Delta^2}}\right)+..,\label{2}\ee
The $\omega^2$ behaviour of $\rho_0(\omega)$ in the neighbourhood of the Fermi level in the small U regime 
is evident in  figure 4 which gives a plot of $\rho_0(\omega)$ for $U/\pi\Delta=0.01$.\par
For $V_0\ne 0$ the zero Majorana fermion is no longer uncoupled from the conduction electrons
and for $U=0$ the corresponding spectral density has a resonance width of $\bar\Delta_0$. 
Having lost the zero mode in this channel the scattering the vector Majorana fermions with
the scalar Majorana fermion with the interaction $U$ is no longer singular. For $\bar\Delta_0/
\bar\Delta\ll
 1$, the resonance in the scalar channel is very narrow 
compared with that of the vector channel, and there is a low energy range in which
the behaviour is very similar to that found in the marginal case. There is a cross-over
eventually on reducing the energy scale  to a Fermi-liquid form of behaviour. This is apparent in
the plots of the spectra in figures 8-10 for the case $\Delta_0/\Delta =10^{-4}$. In figure 8 the spectral
 densities of both the scalar and vector Majorana fermions are shown for the same set
of values for $U/\pi\Delta$ as in figure 1. Due to the small value which was used for the ratio
 $\Delta_0/\Delta$
 the spectra for the vector Majorana fermions shown in figure 9 are quite similar 
to the corresponding curves for the marginal Fermi-liquid case shown in figure 3.  The difference
is only revealed when we compare the insets in the two figures. On the very low energy scale
around the Fermi level there is a distinct flattening and rounding of the spectrum seen in the inset 
in figure 9,
consistent with an $\omega^2$ behaviour in contrast to the  form $|\omega|$ seen in the inset
in figure 3 (which  persisted down to the lowest energy scales for which the spectra were calculated). \par
If we apply second order perturbation theory for finite $\Delta_0$ then, in the  low frequency range
 $\omega\ll \Delta$, the imaginary part of the self-energy 

 has the form,
\be {\rm Im}\Sigma(\omega)=\left({U\over{\pi\Delta}}\right)^2\left[-\omega {\rm tan}^{-1}
\left({\omega\over \Delta_0}\right)+{\Delta_0\over 2}{\rm ln}\left(1+{\omega^2\over \Delta_0^2}
\right)\right].\ee
In the range $\omega>\Delta_0$ this behaves essentially as  $|\omega|$, as in the non-Fermi liquid case 
(\ref{rhovec}). Only in the very low frequency range $\omega\ll \Delta_0$ do we get the influence
of the Fermi-liquid fixed point and the $\omega^2$ behaviour. In the low frequency
 limit $\omega\ll \Delta_0$, the spectral density  of the vector Majorana fermions is then given by
\be \rho_{\alpha}(\omega)={1\over{\pi\Delta}}\left[1-{\omega^2\over{2\Delta\Delta_0}}\left\{{2\Delta_0
\over\Delta}+\left(1+{4\Delta_0\over \Delta}\right)\left({ U\over{\pi\Delta}}\right)^2\right\}\right],\ee
for $\alpha=1,2,3$. For large values of $U$ we can apply the renormalized perturbation approach
and the corresponding result for the spectrum of the vector Majorana fermions in the very low
 frequency limit $\omega\ll \bar\Delta_0$ is 
\be \rho_{\alpha}(\omega)={1\over{\pi\Delta}}\left[1-{\omega^2\over{2\bar\Delta\bar\Delta_0}}
\left\{{2\bar\Delta_0\over\bar\Delta}+\left({\bar U\over{\pi\bar\Delta}}\right)^2\right\}\right].\ee
The spectra for the scalar Majorana fermion for $\Delta_0\ne 0$, however, are very different from those
 for $\Delta_0=0$ as can been seen by comparing the results given in figure 10 with those
in figure 2. There is  no zero mode and the peaks at the Fermi level coincide. In the very low frequency
 range $\omega\ll \bar \Delta_0$ we find for $\rho_0(\omega)$
using the renormalized
perturbation theory the result,

\be \rho_{0}(\omega)={1\over{\pi\Delta_0}}\left[1-{\omega^2\over{2\bar\Delta_0^2}}\left\{2+{{\bar\Delta_0}
\over\bar\Delta}\left({\bar U\over{\pi\bar\Delta}}\right)^2\right\}\right].\ee

The value of the density of states at the Fermi level for the results shown in figure 10  are within
 2\% of the
Freidel sum rule result $1/\pi\Delta_0$ for all values of $U$. The $\omega^2$ behaviour in
the very low energy region near the Fermi-level is clear in the inset in figure 10. 
\subsection{Dynamic spin and charge susceptibilities}\par
\bigskip
We use the same approach to calculate the two particle Green's functions
corresponding to the dynamic spin and charge susceptibilities.
The corresponding spectral densities are of the form
\begin{eqnarray}
 A_Q(\omega)&=&\frac{1}{Z}\sum_{nm}  
           \bigg\vert   \Big< n \Big\vert Q
                                         \Big\vert m \Big>
                   \bigg\vert^2
                   \delta \big( \omega -(E_{n} -E_{m}) \big)
         \nonumber \\ 
         &\quad & \times \left( e^{-\beta E_m} - e^{-\beta E_n} \right)
\end{eqnarray} 
with $Q=g\mu_{\rm B}(n_{d,\uparrow}-n_{d,\downarrow})/2$ for spin
and $Q=(n_{d,\uparrow}+n_{d,\downarrow}-1)/2$ for charge. 
This  is similar in form to that for the single particle Green's function given in 
equation (\ref{specd}), which corresponds  
 to $Q=c^\dagger_{d\sigma}$ with a difference in sign in the factor in the last bracket due
to the use of the commutator  rather than the anticommutator for the spin and charge
Green's functions.
Using the procedure outlined in the appendix these can be expressed in terms of reduced
matrix elements and the degeneracy with respect to $T_z$ can be explicitly summed over (see
equation (\ref{wigner})).
Both the operators for spin and charge transform as irreducible tensors of the form $Q^1_0$,
and the result in each case gives
\begin{eqnarray} 
A_Q(\omega)&=&\frac{1}{3Z}\sum_{nm}  
           \bigg\vert   \Big< n \Big\vert\Big\vert Q
                                       \Big\vert \Big\vert m \Big>
                   \bigg\vert^2
                   \delta \big( \omega -(E_{n} -E_{m}) \big) 
      \nonumber \\
       &\quad &
    \left( e^{-\beta E_m} - e^{-\beta E_n} \right)
\end{eqnarray}
 where $\Big< n \Big\vert\Big\vert Q\Big\vert \Big\vert m \Big>$ is the reduced matrix 
element (see equation (\ref{wigner})).\par
The dynamic susceptibilities for the non-interacting model ($U=0$) can be evaluated analytically.
There are two parts to the d contribution to the spin susceptibility $\chi_{\rm s}(\omega)$. One arises
 from the bubble diagram, figure 11 (a), where  the propagators involved are those 
corresponding to a component of the vector Majorana fermion which we denote by $\chi^{\rm vv}_{\rm d}
(\omega)$ (in which we absorb factors of $g\mu_{\rm B}$).  This contribution is just half
of that obtained for the standard symmetric Anderson model with the real part given by

\be {\rm Re}\chi^{\rm vv}_{\rm d}={{\Delta/4\pi}\over{(\omega^2+4\Delta^2)}}\left\{
{\rm ln}\left|{\Delta^2\over{\omega^2+\Delta^2}}\right|+{4\Delta\over\omega}{\rm tan}^{-1}\left(
{\omega\over\Delta}\right)\right\},\label{rchivv}\ee
and the imaginary part by
\be {\rm Im}\chi^{\rm vv}_{d}(\omega)={{\Delta/2\pi}\over{(\omega^2+4\Delta^2)}}\left\{{\Delta\over\omega}
{\rm ln}\left(1+{\omega^2\over \Delta^2}\right)+{\rm tan}^{-1}\left({\omega\over\Delta}\right)\right\}.
\label{ichivv}\ee
The other contribution, which we denote by $\chi^{\rm vs}_{\rm d}(\omega)$, is from  bubble diagram,
 figure 11(b), in which one of the propagators corresponds to
a component of the vector Majorana fermion while the other propagator is for the scalar
Majorana fermion. The real part is 
given by
\begin{eqnarray}
 {\rm Re}\chi^{\rm vs}_{\rm
d}(\omega)&=&{{\omega^2\Delta/4\pi}\over{(\omega^2+\Delta^2_0-\Delta^2)^2
+4\omega^2\Delta^2}}
\nonumber \\
&\times& 
\left\{\left({{\omega^2-\Delta_0^2+\Delta^2}
\over 
{2\omega^2}}\right){\rm ln}\left|{{\Delta^2}\over
{\omega^2+\Delta_0^2}}\right| \right. \nonumber \\&+&
 \left. {{2\Delta_0}\over{\omega}}{\rm tan}^{-1}
  \left(\omega\over\Delta_0\right)\right\} \nonumber \\&+&
\{{\rm similar\,\, term\,\,with\,\,} \Delta_0\leftrightarrow  
\Delta\},\label{rchivs}
\end{eqnarray}
and the imaginary part by
\begin{eqnarray}
{\rm Im}\chi^{\rm vs}_{\rm d}(\omega)&=&
{{\omega\Delta\Delta_0/4\pi}\over{(\omega^2+\Delta^2_0-
\Delta^2)^2+4\omega^2\Delta^2}} \nonumber \\
&\times&
\left\{\left({{\omega^2-\Delta_0^2+\Delta^2}\over 
{\omega\Delta_0}}\right){\rm tan}^{-1}\left(
\omega\over\Delta_0\right) \right.
\nonumber \\ &+&
  {\rm ln}\left|{(\omega^2+\Delta^2)}\over{\Delta_0^2}\right|
\nonumber \\ &+& {\rm similar\,\, terms\,\, with\,\, } 
\Delta_0\leftrightarrow  \Delta \Bigg\},\label{ichivs}
\end{eqnarray}
In the limit $\Delta_0\to \Delta$ the vector-scalar contribution is the same as the
vector-vector contribution and we recover the results for the dynamic susceptibility
of the symmetric Anderson model. In the limit $\Delta_0\to 0$, where we have a zero mode, 
the real part of the vector-scalar contribution develops a logarithmic singularity,
\be{\rm Re}\chi^{\rm vs}_{\rm d}(\omega)={{\Delta}\over{4\pi(\omega^2+\Delta^2)}}
{\rm ln}\left|{{\Delta}
\over{\omega}}\right|,\label{realpt}\ee
and the imaginary part has a discontinuity at the origin,
\be {\rm Im}\chi^{\rm vs}_d(\omega)={{{\rm sgn}(\omega)\Delta}\over {8(\omega^2+\Delta^2)}}.\label{impt}\ee
We also have some exact reults for $U\ne 0$ from the Shiba relation which relates the imaginary part of
 $\chi_{\rm s}(\omega)/\omega$
for the impurity to the real part of $\chi_{\rm s}(\omega)$ in the limit $\omega\to 0$,
\be {\rm lim}_{\omega\to 0}\left\{{{{\rm Im}\chi_{\rm s}(\omega +i\delta)}\over {2\pi\omega}}\right\}=
{{({\rm Re}\chi_{\rm s}(0))^2}\over {(g\mu_{\rm B})^2}}.\label{shiba}\ee
This result, however, is only applicable to the standard symmetric Anderson model where
 $\Delta_0=\Delta$. \par
We can use the exact results we have for this model to gauge the accuracy of
our NRG calculations. In figure 12 we plot the imaginary part of $\chi_s(\omega)$ divided by the
 square of the real part for various values of $U$. According to the Shiba relation these should all pass
through the same point at $\omega=0$ and this is clearly seen in the results figure 12. The actual
  value at this point is about 8\% too high, reflecting errors of a few \% in the individual 
susceptibilities. The overall accuracy over the full frequency range can be gauged by a comparison
 of the NRG results with the exact results deduced from equations (\ref{rchivv}), (\ref{ichivv}), 
(\ref{rchivs}) and (\ref{ichivs}) for $U=0$. In figure 13 we plot the NRG
results for the real part of $\chi_{\rm s}(\omega)$ for various values of $\Delta_0$, and compare 
them with the analytic results. It can be seen that the two sets of results are in good agreement
 over the
complete frequency range.  \par
To see how the dynamic spin susceptibility for finite $U$ differs from that for the
standard symmetric Anderson model we have plotted both the imaginary part and real part 
in figures 14 and 15 for $U/\pi\Delta=1.5$ and various values of $\Delta_0$. As we reduce $\Delta_0$ 
the imaginary part changes more and more  rapidly at $\omega=0$ until finally
for $\Delta_0=0$ it has a finite discontinuous jump at this point as can be in figure 16, where the 
results for $\Delta_0=0$ are shown for various values of $U$.
The peak in real part of the spin susceptibility narrows considerably on reducing
$\Delta_0$ as can be seen in figure 15. To be able to compare the results for various values of $\Delta_0$
the values shown in figure 15 have been  multiplied by the factor $1/\chi_{\rm s}(0)$,
so it is not so apparent that the narrowing is accompanied by a sharp increase in the absolute
value at low frequencies. In the limit $\Delta_0=0$ the real part of $\chi_{\rm s}(\omega)$ has a
 singularity at $\omega =0$. A discontinuity in the imaginary part of $\chi_s(\omega)$ for $\Delta_0=0$ 
at $\omega=0$ with a logarithmic singularity in the real part is predicted from
equations (\ref{impt}) and (\ref{realpt}) for $U=0$, so it is not surprising that these features
persist for small values of $U$. For large $U$ we can make predictions from our results about
the behaviour of the spin susceptibility of the corresponding two channel Kondo model over the full 
frequency range. We look at this in detail in the next section.\par

\section{Spin dynamics of the two channel Kondo model}\par
In the large $U$ regime we can apply  the mapping  
the O(3) Anderson model in the localized limit and the NRG results of 
previous section to deduce the spin dynamics of the two
channel Kondo model over the complete frequency range. As the mapping is for the spin degrees of freedom
 of the TCKM, which is the only part with an interaction term,  it cannot be used to calculate
all the relevant properties of the TCKM, only those where the other degrees of freedom of the conduction
electrons do not
contribute.  It can be used for example to deduce the  the impurity contribution to the thermodyamics,
such as susceptibility and specific heat, and  also  for the spin dynamics of the impurity, where the
 uncoupled degrees of freedom  factorize and cancel out.
This is a distinct advantage 
for NRG calculations to use the equivalent  in one channel model because a  larger percentage of the states
can be retained after each diagonalization in the iterative procedure. 
The number of states that can retained be at each stage 
is proportional to $1/4^{N_c}$, where $N_c$ is the number of channels, so this can be serious
limitation when the number of channels is increased. However, quite apart from these practical
considerations, 
the relation between the TCKM and  the
 the O(3) Anderson model provides
a simple way of understanding in terms of renormalized Majorana fermions \cite{cit,bhz}. \par
We saw in the previous section that the real part of the dynamic spin susceptibility ($\Delta_0=0$)
at $T=0$ has a logarithmic singularity in $\omega$ for $U=0$ and that in the NRG results this singular
 behaviour persists for finite
$U$. Such a term in ln$\omega$ has also been found in direct NRG calculations of the dynamic susceptibility
 for the TCKM \cite{shi}.  To analyse the situation in the large $U$ limit
 to use the mapping to the TCKM it is useful to work with a
 conserved quantity, which in this case is not the spin but ${\bf T}$ the spin plus isospin.
The corresponding local dynamic susceptibility we denote by $\chi_{\bf t}(\omega)$, where
${\bf t}={\bf \sigma}_d+{\bf \tau}_d$. Because the charge fluctuations of the impurity are
suppressed for large $U$ this susceptibility becomes identical with the spin susceptibility
in the large $U$ limit. We can use the NRG approach described in the last section to calculate
 $\chi_{\bf t}(\omega)$ and verify this explicitly. We find for $U/\pi\Delta=4$ at the most a 5\%
 difference
between $\chi_{\bf t}(\omega)$ and  $\chi_{\rm s}(\omega)$ over the complete low  frequency range
$\omega\ll U$.\par
For the large $U$ regime we can  apply our renormalized perturbation approach to calculate
the $\omega$ dependence of $\chi_{\bf t}(\omega)$, following the same steps we used in our earlier
 work to calculate 
its temperature dependence in the zero frequency limit \cite{bhz}. We first calculate the contribution
due to the non-interacting renormalized Majorana fermions. This is given by the vector-vector bubble
 shown in figure 11 (a), together with an overall factor of 4 and $\Delta$ replaced by $\bar\Delta$
so that for $T=0$, 
\be{\rm Re}\chi_{\bf t}(\omega)={{\bar\Delta}\over{\pi(\omega^2+4\bar\Delta^2)}}\left\{
{\rm ln}\left|{\bar\Delta^2\over{\omega^2+\bar\Delta^2}}\right|+{4\bar\Delta\over\omega}{\rm tan}^{-1}
\left(
{\omega\over\bar\Delta}\right)\right\}\label{chdyn}.\ee
Because the single particle Green's function and the quasiparticle Green's function 
in the renormalized perturbation approach differ by a wavefunction renormalization factor $\bar z$,
one might have expected an extra $\bar z^2$ multiplying (\ref{chdyn}) to give the two particle
Green's function corresponding to the dynamic susceptibility. This is not the case because
there is a compensating factor associated with the coupling to the external field.  
The net result is that these two terms cancel so that in the quasiparticle Hamiltonian in the presence of a
 external field the quasi-particles are coupled to the field with the same coupling
as the bare particles, i.e. there is no g-factor renormalization.  
In the zero frequency limit (\ref{chdyn}) gives $\chi_{\bf t}(0)=1/\pi\bar\Delta$, which corresponds
to zero temperature impurity contribution to the total susceptibility we found in our earlier
work in the same limit. The leading order correction term to this result  arises from 
a second order term in $\bar U$, shown in figure 17, where the scalar-vector bubble shown in figure 11, 
 gives a ln$\omega$ contribution similar to the ln$T$ dependence in the same susceptibility at $\omega=0$
 see \cite {bhz}, so that asymptotically as $\omega\to 0$ we find
\be {\rm Re}\chi_{\bf t}(\omega)=\left({1\over {\pi\bar\Delta}}- 
\left ({\bar U\over {\pi\bar\Delta}}\right)^2{{{\rm ln}(\omega/\bar\Delta)}
\over{\pi\bar\Delta}}\right).\label{ca}\ee
The correspondence of the calculation of the ln$\omega$ terms with the calculation of the ln$T$ terms for
  the zero frequency susceptibility is so precise that we can appeal to arguments
 in our earlier work \cite{bhz} to show that  higher order ln$\omega$ terms 
cancel in the same way as the higher order ln$T$ terms. \par
Though we cannot explicitly carry out the full renormalization programme to calculate the renormalized
parameters $\bar U$ and $\bar\Delta$ in terms of the bare parameters $U$ and $\Delta$,
we can use the results $\bar U/\pi\bar\Delta=1$, $\pi\bar\Delta=T_{\rm K}$ from our earlier
work, which were  obtained by equating the expressions for the temperature dependence of the
susceptibility in the large $U$ limit. We then find for the asymptotic behaviour of
the dynamic susceptibility of the two channel Kondo model $\chi_{\rm KM}(\omega)$ as $\omega\to 0$,
$\omega=0$ see\cite {bhz}, so that asymptotically as $\omega\to 0$ we find
\be {\rm Re}\chi_{\rm KM}(\omega)={1\over T_{\rm K}}\left(1- 
{\rm ln}(\omega\pi/T_{\rm K})
\right).\label{cb}\ee
In figure 18 we plot the real part of both $\chi_{\bf t}(\omega)$ and $\chi_{\rm s}(\omega)$ versus
 ${\rm ln}(\omega)$ for $U/\pi\Delta=4$ and the asymptotic form given in equation (\ref{cb}) is 
clearly seen in the low frequency limit. We have also calculated $T_{\rm K}$ both from the limiting
 form  of $\chi_{\bf t}(T)$
as $ T\to 0$ and $\chi_{\bf t}(\omega)$
as $ \omega\to 0$, for $U/\pi\Delta=4$ and $V=0.00141$ and find $T_{\rm K}=1.3\pm 0.2$ and 
 $T_{\rm K}=1.28\pm 0.1$, respectively, which confirms the result 
(\ref{cb}) to within the error limits of the calculations. \par

\section{Conclusions}
We have shown that O(3) Anderson model in the large $U$ limit can describe precisely
the interacting spin degrees of freedom of the linear dispersion two channel Kondo model. 
This extends earlier work in that it establishes that this relationship applies over the full energy
 range of the two channel model
and not just in the immediate region of the low energy fixed point. It enables us to
calculate the impurity contribution to the thermodynamics, and also the impurity spin
dynamics, for the TCKM from the O(3) Anderson model in the strong correlation regime. 
This has a number of advantages. The O(3) model is most conveniently expressed in terms
of Majorana fermions and the low energy fixed point of this model corresponds to free
Majorana fermions whatever the value of the on-site interaction $U$, so the nature of
the fixed point is the same in weak and strong coupling. This is similar to the
standard Anderson model, which has a Fermi liquid fixed point, independent of the
value of $U$. The main difference between the O(3) and the standard Anderson model,
is that O(3) model has a zero mode in the case that corresponds to the channel isotropic
TCKM which results in singular scattering of the renormalized Majorana fermions giving
rise to logarithmically divergent contributions
to the impurity specific heat coefficient, susceptibility, and dynamic spin-spin response
function. This singular low energy behaviour corresponds the marginal Fermi liquid theory
 that was put forward to describe the anomalous behaviour observed
in the normal state of the cuprate superconductors \cite{marg}. 
 The description  of the TCKM in terms of Majorana
fermions not only gives a simple interpretation of the many-body excitations at the low energy
fixed point, both for the channel isotropic and
anisotropic cases, but can also be applied on all energy scales.\par
To find the renormalizations of the Majorana fermions explicitly we use the NRG approach
to calculate the Majorana  spectral densities.
We have calculated weight of the zero Majorana fermion mode and have shown that it decreases
exponentially with $U$
in the large $U$ regime. We have also demonstrated the $|\omega|$ behaviour of
the other Majorana fermion modes in the marginal Fermi liquid case as $\omega\to 0$.
We have used the same approach to calculate the two particle response functions
which we have determined over the full frequency range for $T=0$. We have exploited the mapping between
 the models to deduce the local dynamic spin susceptibility of the TCKM. Renormalized perturbation theory
 has been used to interpret the results and to calculate 
the coefficient of the ln$\omega$ term in asymptotic form of the real part of the dynamic spin
 susceptibility of the TCKM in the limit $\omega\to 0$. \par

\bigskip
\noindent{\bf Acknowledgment}\par
\bigskip
We are grateful  for the support of an EPSRC research grant (GR/J85349), and for a DFG  research grant 
(Bu965-1/1) for one of us (RB).\par

\section{Appendix}\par
Here we outline the steps involved in  the evaluation of the spectral density  given in equation
(\ref{specd}).
The many-body states $|n\rangle$ are classified using the quantum numbers $T$ and $T_z$
of the total angular momentum (spin plus isospin) and expressed in the form
 $|T_n,T_{z,n},w_n\rangle$, where  $w_n$ labels any further degeneracies of the state  $|n\rangle$.
The degeneracies associated with the z-component of angular momentum in (\ref{specd}) 
can be summed over
explicitly. The creation and annihilation operators,
 $c^{\dagger}_{d,\sigma}$ and $c^{\dagger}_{d,\sigma}$, first of all
have to be expressed in terms of the irreducible tensor operators $V_q^k$, k=1,0, q=1,0,-1.
From equation (32) in  reference \cite{bh},
$$V^0_0={1\over{\sqrt{2}}}(c^{}_{d,\downarrow}+c^{\dagger}_{d,\downarrow})$$
$$V^1_1=c^{\dagger}_{d,\uparrow}$$
$$V^1_0={1\over{\sqrt{2}}}(c^{}_{d,\downarrow}-c^{\dagger}_{d,\downarrow})$$
\be V^1_{-1}=-c^{}_{d,\uparrow}\ee
Then the  Wigner-Eckart Theorem,
$$ \langle T',T'_z,w'|V^k_q|T,T_z,w\rangle={1\over{\sqrt{2T'+1}}}
\langle T',w'||V^k_q||T,w\rangle$$
\be \times \langle T,T_z,k,q|T',T'_z,\rangle\label{wigner}\ee
is used to express all the matrix elements in terms of the reduced matrix elements 
$\langle T',w'||V^k_q||T,w\rangle$  which are independent of 
 the $T_z$ quantum numbers, where $\langle T,T_z,k,q|T',T'_z,\rangle$ are the
 Clebsch-Gordon coefficients.
For example for the spin down matrix elements we find
\begin{eqnarray}
           \bigg\vert   \Big< n \Big\vert c^\dagger_{d\downarrow}
                          \Big\vert m \Big> \bigg\vert^2 
   & = & \bigg\vert   \Big< T_n,T_{z,n}, w_n  \Big\vert c^\dagger_{d\downarrow}
                          \Big\vert T_m,T_{z,m}, w_m \Big> \bigg\vert^2 
       \nonumber\\
  & = & \bigg\vert   \Big< T_n,T_{z,n}, w_n  \Big\vert 
          \frac{1}{\sqrt{2}} \left( V_{0}^1 - V_{0}^0 \right)
                          \Big\vert T_m,T_{z,m}, w_m \Big> \bigg\vert^2 
          \nonumber\\
  & = & \frac{1}{2(2T_n+1)}  \bigg\vert  
      \Big< T_n, w_n  \Big\vert\Big\vert 
           V_{0}^1  \Big\vert  \Big\vert T_m, w_m \Big> 
        \Big<T_m,T_{z,m},1,0 \Big\vert T_n,T_{z,n}\Big> 
           \nonumber\\
    & &    - \ \Big< T_n, w_n  \Big\vert\Big\vert 
           V_{0}^0  \Big\vert  \Big\vert T_m, w_m \Big> 
           \delta_{T_n,T_m}\delta_{T_{z,n},T_{z,m}}
\bigg\vert^2 
\end{eqnarray}
For $T_n \ne T_m$, the $V_{0}^0$-term does not contribute and for 
this term we find
\begin{eqnarray}
\sum_{T_{z}}   
           \bigg\vert   \Big< n \Big\vert c^\dagger_{d\downarrow}
                          \Big\vert m \Big> \bigg\vert^2 
  &=& \frac{1}{2(2T_n+1)} \bigg\vert  
      \Big< T_n, w_n  \Big\vert\Big\vert 
           V_{0}^1  \Big\vert  \Big\vert T_m, w_m \Big> \bigg\vert^2
     \sum_{T_{z}} \bigg\vert  
        \Big<T_m,T_{z},1,0 \Big\vert T_n,T_{z}\Big> \bigg\vert^2
 \nonumber \\
  &=& \frac{1}{6} \bigg\vert  
      \Big< T_n, w_n  \Big\vert\Big\vert 
           V_{0}^1  \Big\vert  \Big\vert T_m, w_m \Big> \bigg\vert^2     
\end{eqnarray}
For the case $T_n\!=\!T_m\!=T$ we find
\begin{equation}
\sum_{T_{z}}   
           \bigg\vert   \Big< n\Big\vert c^\dagger_{d\downarrow}
                          \Big\vert m\Big> \bigg\vert^2 
  = \frac{1}{6}  \bigg\vert\Big< T, w_n  \Big\vert\Big\vert 
           V_{0}^1  \Big\vert  \Big\vert T, w_m \Big> \bigg\vert^2
+ \frac{1}{2}  \bigg\vert\Big< T, w_n  \Big\vert\Big\vert 
           V_{0}^0  \Big\vert  \Big\vert T, w_m \Big> \bigg\vert^2
\end{equation}

The reduced matrix elements $\Big< T_n, w_n  \Big\vert\Big\vert 
           V_{0}^{q}  \Big\vert  \Big\vert T_m, w_m \Big> $
are calculated from those of the previous iteration in two steps.
We start from the expression for the reduced matrix element with 
respect to the eigenstates of the impurity-conduction electron
chain with $N+1$ sites,
\begin{eqnarray}
  \Big< T_n, w_n  \Big\vert\Big\vert 
           V_{0}^{q}  \Big\vert  \Big\vert T_m, w_m \Big>_{N+1} 
 = \sum_{r_n,p_n,r_m,p_m} U_{T_n}(w_n,r_n,p_n) U_{T_m}(w_m,r_m,p_m)
     \nonumber \\ \times
      \Big< T_n, r_n,p_n \Big\vert\Big\vert 
           V_{0}^{q}  \Big\vert  \Big\vert T_m, r_m,p_m \Big>_{N+1},
       \label{eq:3mat}\end{eqnarray}
 in terms of the basis states 
$\Big\vert T_n, r_n,p_n \Big>_{N+1}$ used for this calculation
(these are constructed from the eigenstates of 
the $N$ site chain $\Big\vert T_n, r_n \Big>_N$  
 and  the states for the additional site labelled by $p_n$, see reference \cite{bh}
for further details). The coefficients $U_{T_n}(w_n,r_n,p_n)$ are the 
components of the eigenstate $ \Big\vert T_n, w_n \Big>_{N+1}$
with respect to this basis. 
The reduced matrix elements $\Big< 
T_n,r_n,p_n \Big\vert\Big\vert 
           V_{0}^{q}  \Big\vert  \Big\vert T_m, r_m,p_m \Big>_{N+1}$
have to be related to the reduced matrix elements of the previous step
as follows:
\begin{eqnarray}
 \Big< T,r,1 \Big\vert\Big\vert V_{0}^{1} \Big\vert\Big\vert
           T+1,r^\prime,1 \Big>_{N+1} &=&
       \sqrt{\frac{2T+3}{2(T+1)}} 
 \Big< T-\frac{1}{2},r\Big\vert\Big\vert V_{0}^{1} \Big\vert\Big\vert
           T+\frac{1}{2},r^\prime\Big>_N \nonumber \\ 
   \Big< T,r,1 \Big\vert\Big\vert V_{0}^{1} \Big\vert\Big\vert
           T-1,r^\prime,1 \Big>_{N+1} &=&
        \sqrt{\frac{2T+1}{2T}}
 \Big< T-\frac{1}{2},r\Big\vert\Big\vert V_{0}^{1} \Big\vert\Big\vert
           T-\frac{3}{2},r^\prime\Big>_N \nonumber \\ 
  \Big< T,r,1 \Big\vert\Big\vert V_{0}^{1} \Big\vert\Big\vert
           T,r^\prime,1 \Big>_{N+1} &=&
       \frac{1}{\sqrt{2}T} \sqrt{(T+1)(2T-1)} 
 \Big< T-\frac{1}{2},r\Big\vert\Big\vert V_{0}^{1} \Big\vert\Big\vert
           T-\frac{1}{2},r^\prime\Big>_N \nonumber \\ 
  \Big< T,r,1 \Big\vert\Big\vert V_{0}^{0} \Big\vert\Big\vert
           T-1,r^\prime,1 \Big>_{N+1} &=&
      \sqrt{\frac{2T+1}{2T}}
  \Big< T-\frac{1}{2},r\Big\vert\Big\vert V_{0}^{0} \Big\vert\Big\vert
           T-\frac{1}{2},r^\prime\Big>_N \nonumber
\end{eqnarray}

\begin{eqnarray}
   \Big< T,r,2 \Big\vert\Big\vert V_{0}^{1} \Big\vert\Big\vert
           T+1,r^\prime,2 \Big>_{N+1} &=&
    -\sqrt{\frac{2T+3}{2(T+1)}}
  \Big< T-\frac{1}{2},r\Big\vert\Big\vert V_{0}^{1} \Big\vert\Big\vert
           T+\frac{1}{2},r^\prime\Big>_N \nonumber \\ 
   \Big< T,r,2 \Big\vert\Big\vert V_{0}^{1} \Big\vert\Big\vert
           T-1,r^\prime,2 \Big>_{N+1} &=& -
        \sqrt{\frac{2T+1}{2T}}
   \Big< T-\frac{1}{2},r\Big\vert\Big\vert V_{0}^{1} \Big\vert\Big\vert
           T-\frac{3}{2},r^\prime\Big>_N \nonumber \\ 
   \Big< T,r,2 \Big\vert\Big\vert V_{0}^{1} \Big\vert\Big\vert
           T,r^\prime,2 \Big>_{N+1} &=& -
         \frac{1}{\sqrt{2}T} \sqrt{(T+1)(2T-1)} 
   \Big< T-\frac{1}{2},r\Big\vert\Big\vert V_{0}^{1} \Big\vert\Big\vert
           T-\frac{1}{2},r^\prime\Big>_N \nonumber \\ 
   \Big< T,r,2 \Big\vert\Big\vert V_{0}^{0} \Big\vert\Big\vert
           T-1,r^\prime,2 \Big>_{N+1} &=& -
       \sqrt{\frac{2T+1}{2T}}
  \Big< T-\frac{1}{2},r\Big\vert\Big\vert V_{0}^{0} \Big\vert\Big\vert
           T-\frac{1}{2},r^\prime\Big>_N \nonumber
\end{eqnarray}

\begin{eqnarray}
   \Big< T,r,3 \Big\vert\Big\vert V_{0}^{1} \Big\vert\Big\vert
          T+1,r^\prime,3 \Big>_{N+1} &=& -
         \sqrt{\frac{2T+1}{2(T+1)}} 
   \Big< T+\frac{1}{2},r\Big\vert\Big\vert V_{0}^{1} \Big\vert\Big\vert
           T+\frac{3}{2},r^\prime\Big>_N \nonumber \\
   \Big< T,r,3 \Big\vert\Big\vert V_{0}^{1} \Big\vert\Big\vert
           T-1,r^\prime,3 \Big>_{N+1} &=& -
  \sqrt{\frac{(T+1)(2T-1)}{2T(T+1)}} 
    \Big< T+\frac{1}{2},r\Big\vert\Big\vert V_{0}^{1} \Big\vert\Big\vert
           T-\frac{1}{2},r^\prime\Big>_N \nonumber \\
   \Big< T,r,3 \Big\vert\Big\vert V_{0}^{1} \Big\vert\Big\vert
           T,r^\prime,3 \Big>_{N+1} &=& -
  \frac{1}{\sqrt{2}(T+1)}\sqrt{T(2T+3)}
   \Big< T+\frac{1}{2},r\Big\vert\Big\vert V_{0}^{1} \Big\vert\Big\vert
           T+\frac{1}{2},r^\prime\Big>_N \nonumber \\
   \Big< T,r,3 \Big\vert\Big\vert V_{0}^{0} \Big\vert\Big\vert
           T,r^\prime,3 \Big>_{N+1} &=& -
               \sqrt{\frac{2T+1}{2(T+1)}} 
  \Big< T+\frac{1}{2},r\Big\vert\Big\vert V_{0}^{0} \Big\vert\Big\vert
           T+\frac{1}{2},r^\prime\Big>_N \nonumber
\end{eqnarray}

\begin{eqnarray}
   \Big< T,r,4 \Big\vert\Big\vert V_{0}^{1} \Big\vert\Big\vert
           T+1,r^\prime,4 \Big>_{N+1} &=& 
         \sqrt{\frac{2T+1}{2(T+1)}} 
   \Big< T+\frac{1}{2},r\Big\vert\Big\vert V_{0}^{1} \Big\vert\Big\vert
           T+\frac{3}{2},r^\prime\Big>_N \nonumber \\
   \Big< T,r,4 \Big\vert\Big\vert V_{0}^{1} \Big\vert\Big\vert
           T-1,r^\prime,4 \Big>_{N+1} &=& 
  \sqrt{\frac{(T+1)(2T-1)}{2T(T+1)}} 
    \Big< T+\frac{1}{2},r\Big\vert\Big\vert V_{0}^{1} \Big\vert\Big\vert
           T-\frac{1}{2},r^\prime\Big>_N \nonumber \\
   \Big< T,r,4 \Big\vert\Big\vert V_{0}^{1} \Big\vert\Big\vert
           T,r^\prime,4 \Big>_{N+1} &=& 
  \frac{1}{\sqrt{2}(T+1)}\sqrt{T(2T+3)}
    \Big< T+\frac{1}{2},r\Big\vert\Big\vert V_{0}^{1} \Big\vert\Big\vert
           T+\frac{1}{2},r^\prime\Big>_N \nonumber \\
   \Big< T,r,4 \Big\vert\Big\vert V_{0}^{0} \Big\vert\Big\vert
           T,r^\prime,4 \Big>_{N+1} &=& 
               \sqrt{\frac{2T+1}{2(T+1)}} 
   \Big< T+\frac{1}{2},r\Big\vert\Big\vert V_{0}^{0} \Big\vert\Big\vert
           T+\frac{1}{2},r^\prime\Big>_N \nonumber
\end{eqnarray}

\begin{eqnarray}
   \Big< T,r,1 \Big\vert\Big\vert V_{0}^{1} \Big\vert\Big\vert
           T,r^\prime,4 \Big>_{N+1} &=& 
      \frac{(-1)^{N+1}}{\sqrt{2T(T+1)}}
   \Big< T-\frac{1}{2},r\Big\vert\Big\vert V_{0}^{1} \Big\vert\Big\vert
           T+\frac{1}{2},r^\prime\Big>_N \nonumber \\
   \Big< T,r,1 \Big\vert\Big\vert V_{0}^{1} \Big\vert\Big\vert
           T-1,r^\prime,4 \Big>_{N+1} &=& 
      \frac{(-1)^{N+1}}{\sqrt{2}T}
  \Big< T-\frac{1}{2},r\Big\vert\Big\vert V_{0}^{1} \Big\vert\Big\vert
           T-\frac{1}{2},r^\prime\Big>_N \nonumber
\end{eqnarray}

\begin{eqnarray}
   \Big< T,r,2 \Big\vert\Big\vert V_{0}^{1} \Big\vert\Big\vert
           T,r^\prime,3 \Big>_{N+1} &=& 
      \frac{1}{\sqrt{2T(T+1)}}
  \Big< T-\frac{1}{2},r\Big\vert\Big\vert V_{0}^{1} \Big\vert\Big\vert
           T+\frac{1}{2},r^\prime\Big>_N  \nonumber \\
   \Big< T,r,2 \Big\vert\Big\vert V_{0}^{1} \Big\vert\Big\vert
           T-1,r^\prime,3 \Big>_{N+1} &=& 
      \frac{1}{\sqrt{2}T}
  \Big< T-\frac{1}{2},r\Big\vert\Big\vert V_{0}^{1} \Big\vert\Big\vert
           T-\frac{1}{2},r^\prime\Big>_N \nonumber
\end{eqnarray}

\begin{eqnarray}
   \Big< T,r,4 \Big\vert\Big\vert V_{0}^{1} \Big\vert\Big\vert
           T+1,r^\prime,1 \Big>_{N+1} &=& 
      \frac{(-1)^{N}}{\sqrt{2}(T+1)}
  \Big< T+\frac{1}{2},r\Big\vert\Big\vert V_{0}^{1} \Big\vert\Big\vert
           T+\frac{1}{2},r^\prime\Big>_N  \nonumber \\
   \Big< T,r,4 \Big\vert\Big\vert V_{0}^{1} \Big\vert\Big\vert
           T,r^\prime,1 \Big>_{N+1} &=& 
      \frac{(-1)^{N}}{\sqrt{2T(T+1)}}
   \Big< T+\frac{1}{2},r\Big\vert\Big\vert V_{0}^{1} \Big\vert\Big\vert
           T-\frac{1}{2},r^\prime\Big>_N \nonumber
\end{eqnarray}

\begin{eqnarray}
   \Big< T,r,3 \Big\vert\Big\vert V_{0}^{1} \Big\vert\Big\vert
           T+1,r^\prime,2 \Big>_{N+1} &=& 
      \frac{-1}{\sqrt{2}(T+1)}
   \Big< T+\frac{1}{2},r\Big\vert\Big\vert V_{0}^{1} \Big\vert\Big\vert
           T+\frac{1}{2},r^\prime\Big>_N  \nonumber\\
   \Big< T,r,3 \Big\vert\Big\vert V_{0}^{1} \Big\vert\Big\vert
           T,r^\prime,2 \Big>_{N+1} &=& 
      \frac{-1}{\sqrt{2T(T+1)}}
   \Big< T+\frac{1}{2},r\Big\vert\Big\vert V_{0}^{1} \Big\vert\Big\vert
           T-\frac{1}{2},r^\prime\Big>_N  \nonumber
\end{eqnarray}

The matrix elements between {\it all} states at intermediate steps
have to be calculated. 
The initial values are
\begin{eqnarray}
    \Big< \frac{1}{2},1  \Big\vert\Big\vert V_{0}^{1}
       \Big\vert\Big\vert \frac{1}{2},2 \Big> &=&
        -\sqrt{3} \nonumber \\
    \Big< \frac{1}{2},1  \Big\vert\Big\vert V_{0}^{0}
       \Big\vert\Big\vert \frac{1}{2},2 \Big> &=&
        1 \nonumber \\
    \Big< \frac{1}{2},2  \Big\vert\Big\vert V_{0}^{1}
       \Big\vert\Big\vert \frac{1}{2},1 \Big> &=&
        -\sqrt{3}  \nonumber \\
    \Big< \frac{1}{2},2  \Big\vert\Big\vert V_{0}^{0}
       \Big\vert\Big\vert \frac{1}{2},1 \Big>&=&
        - 1 \nonumber
\end{eqnarray}

The spectral density of the spin up Green's function can be evaluated in a similar
way and is found to depend only on the matrix element of the tensor operator $V^1_0$
and not on $V_0^0$.
This Green's function is the same as that of the $\alpha=1$
or $\alpha=2$ Majorana fermion Green's functions, which are equal and equivalent to the
Green's function for the
$\alpha=3$ Majorana fermion as a result of the O(3) symmetry of the model. 
We can deduce that $V^1_0$ is associated with the $\alpha=1,2,3$
Majorana fermions and  $V^0_0$  with the $\alpha=0$ Majorana fermion.
As spin down Green's function is the half the sum of the Green's functions
for the Majorana fermions for $\alpha=0,3$ we can deduce that the term involving
$V_0^0$ in the final formula for the spectral density of the spin down Green's function
is   the spectral density of the $\alpha=0$ Majorana fermion (apart from a factor 1/2)
and, similarly, the term in $V_0^1$ is the spectral density of the $\alpha=1,2,3$
Majorana fermions.
%
%
%
\begin{table}
\caption{Many-body excitation energies and their
  corresponding degeneracies ($dg$) of the two channel Kondo model
corresponding to the single particle spectrum at the low-energy 
fixed point. The energy levels without primes correspond to sector A, 
and those with primes to sector B. 
}
\label{tab:1}       
\begin{tabular}[t]{cccc}
   $E_{\rm ex}/(\pi v_{\rm F}/l)$  & $\sum_k n_k \epsilon_k $
       & $dg$   & total $dg$\\
  \hline
   0 & 0 & 2 & 2 \\
  \hline
   1/8 & $0'$ & 4 & 4 \\
   \hline
   1/2 &$\varepsilon_{1/2}$ & 10 & 10 \\
  \hline
  5/8  &$\varepsilon'_{1/2}$ & 12 & 12 \\
  \hline
   1 &$\varepsilon_{1}$ & 6 &  \\
     &$2\varepsilon_{1/2}$ & 20 & 26 \\
  \hline
  9/8 &$\varepsilon'_{1}$ & 20 &  \\
     &$2\varepsilon'_{1/2}$ & 12 & 32 \\
  \hline
   3/2 &$\varepsilon_{3/2}$ & 10 &  \\
       &$\varepsilon_{1} + \varepsilon_{1/2}$ & 30 & 60 \\
       &$3\varepsilon_{1/2}$ & 20 &  \\
   \hline
   13/8 &$\varepsilon'_{3/2}$ & 12 &  \\
       &$\varepsilon'_{1} + \varepsilon'_{1/2}$ & 60 & 76 \\
\end{tabular}
\end{table}

%
%

\newpage

\begin{center}

\begin{figure}
\epsfxsize=5.0in
\hspace*{0.0cm}\epsffile{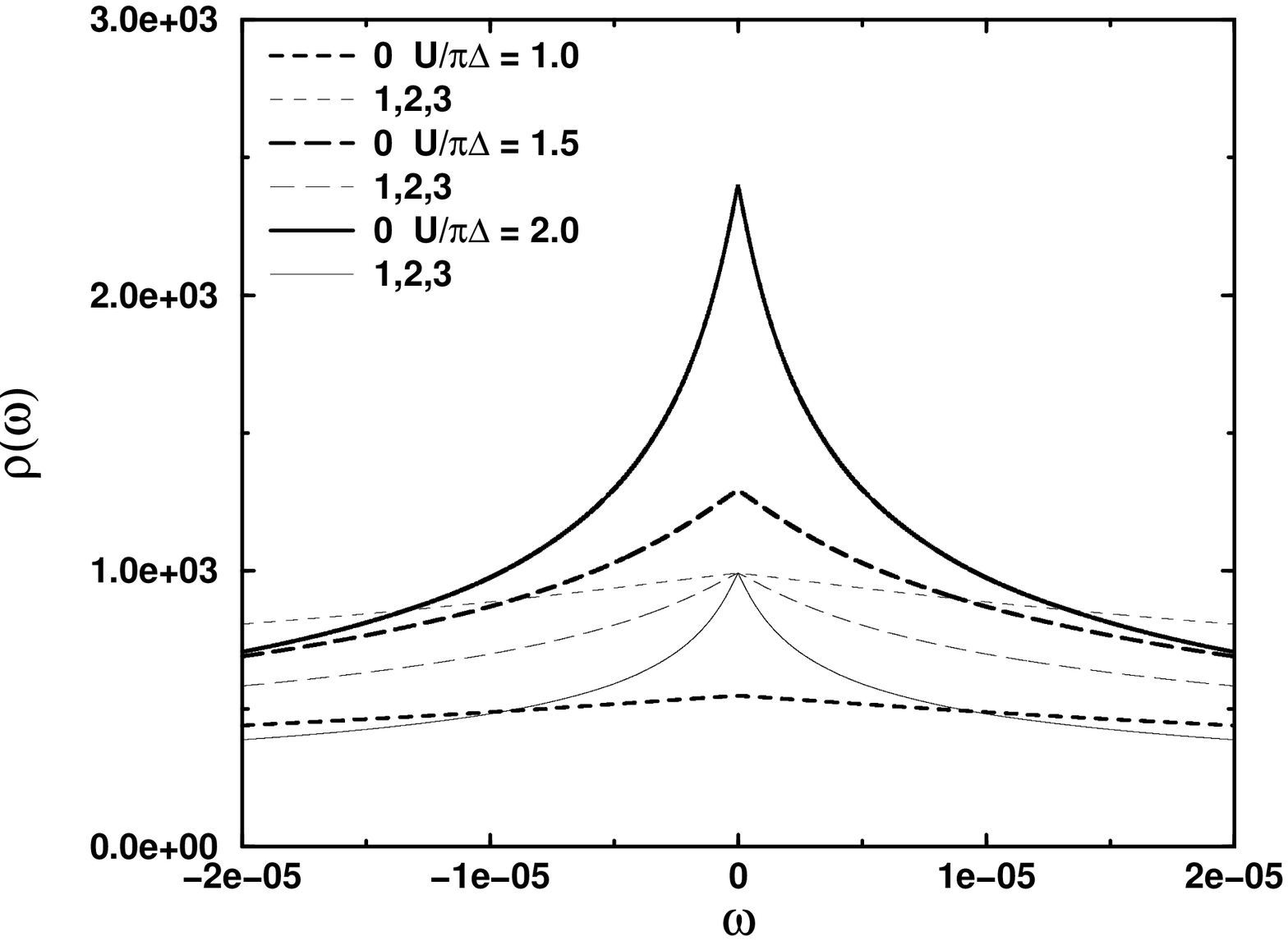}
\caption{The spectral density 
\( \rho \left( \omega \right) \) of the Majorana 
fermion impurity Green's function in the marginal
 Fermi liquid situation: \( V=0.01414, V_0=0.0 \  
\left( \Delta = 10^{-4}\pi, \Delta_0 = 0.0 \right) \).
}
\label{fig:1}       
\end{figure}

\begin{figure}
\epsfxsize=5.0in
\hspace*{0.0cm}\epsffile{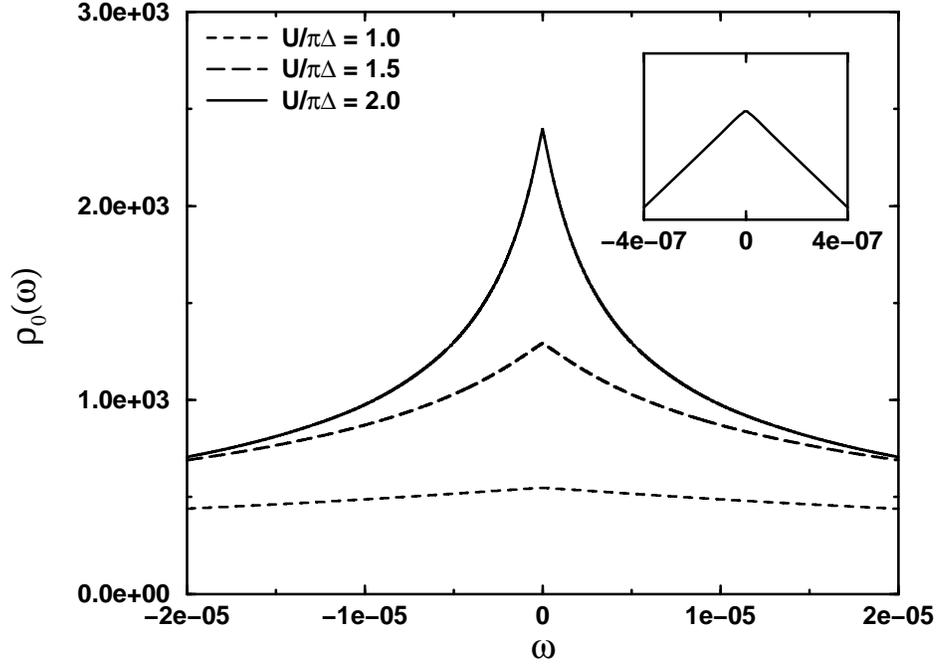}
\caption{The spectral density \( \rho_{0} \left( 
\omega \right) \) of the \( \alpha=0\) Majorana 
fermion Green's function in the marginal Fermi 
liquid situation: \( V=0.01414, V_0=0.0 \ \left( 
\Delta = 10^{-4}\pi, \Delta_0 = 0.0 \right) \). 
The inset shows the \( |\omega| \) dependence 
near the Fermi level.}
\label{fig:2}       
\end{figure}

\begin{figure}
\epsfxsize=5.0in
\hspace*{0.0cm}\epsffile{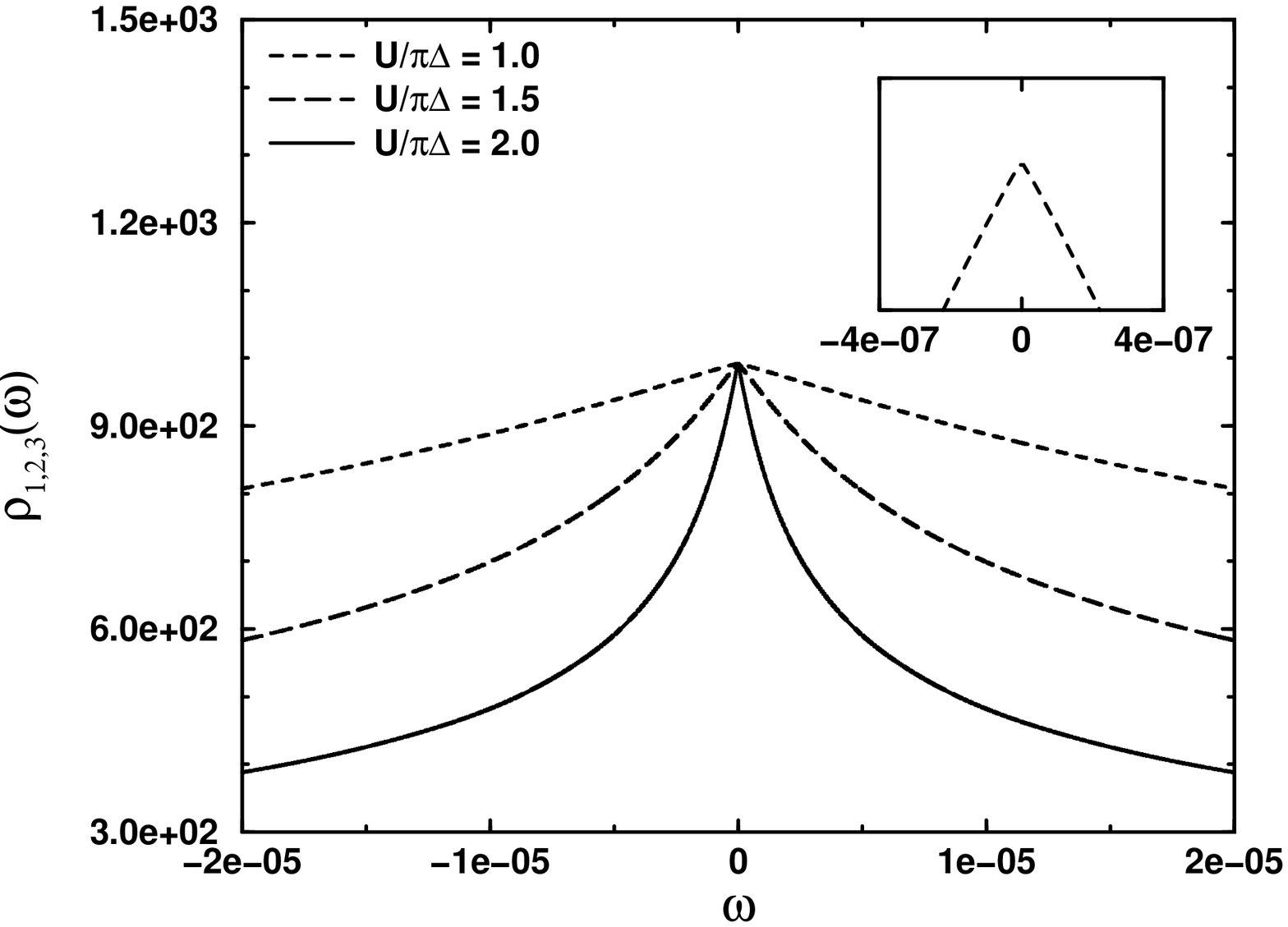}
\caption{The spectral density \( \rho_{1,2,3} \left( 
\omega \right) \) of the \( \alpha=1,2,3 \) Majorana 
fermion Green's function in the marginal Fermi liquid 
situation: \( V=0.01414, V_0=0.0 \ \) \( \left( 
\Delta = 10^{-4}\pi, \Delta_0 = 0.0 \right)\). 
The inset shows the \( |\omega| \) dependence near 
the Fermi level. \( \rho_{1,2,3} \left( 0 \right) 
\) is within 2\% of the value given by the Friedel sum rule.}
\label{fig:3}       
\end{figure}

\begin{figure}
\epsfxsize=5.0in
\hspace*{0.0cm}\epsffile{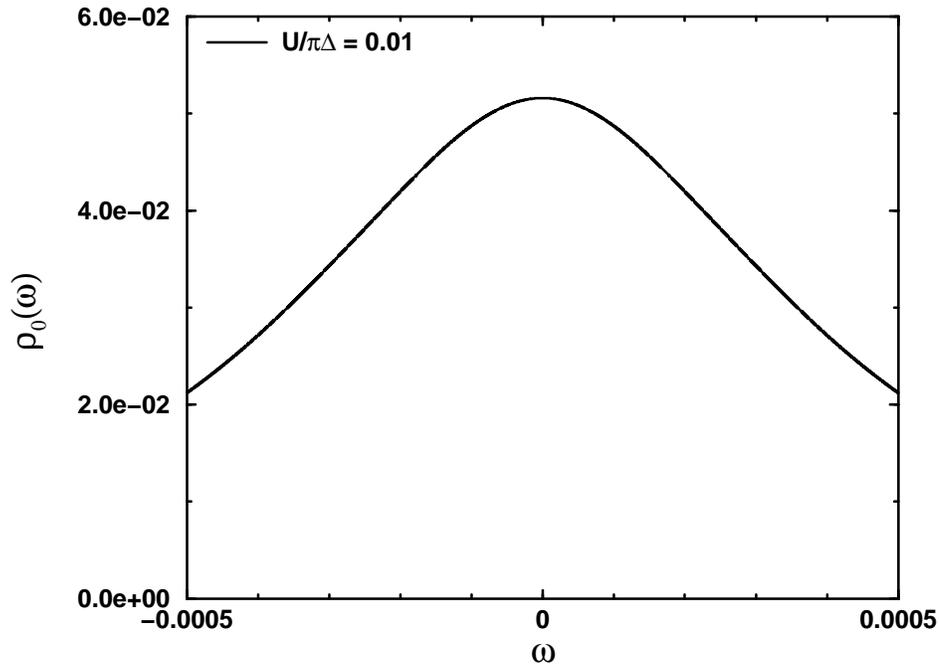}
\caption{The spectral density \( \rho_{0} \left( 
\omega \right) \) of the \( \alpha=0\) Majorana 
fermion Green's function in the marginal Fermi 
liquid situation: \( V=0.01414, V_0=0.0 \ 
\left( \Delta = 10^{-4}\pi, \Delta_0 = 
0.0 \right) \), indicating the influence 
of \(w^2\) corrections for small \(U/\pi\Delta\).}
\label{fig:4}       
\end{figure}

\begin{figure}
\epsfxsize=5.0in
\hspace*{0.0cm}\epsffile{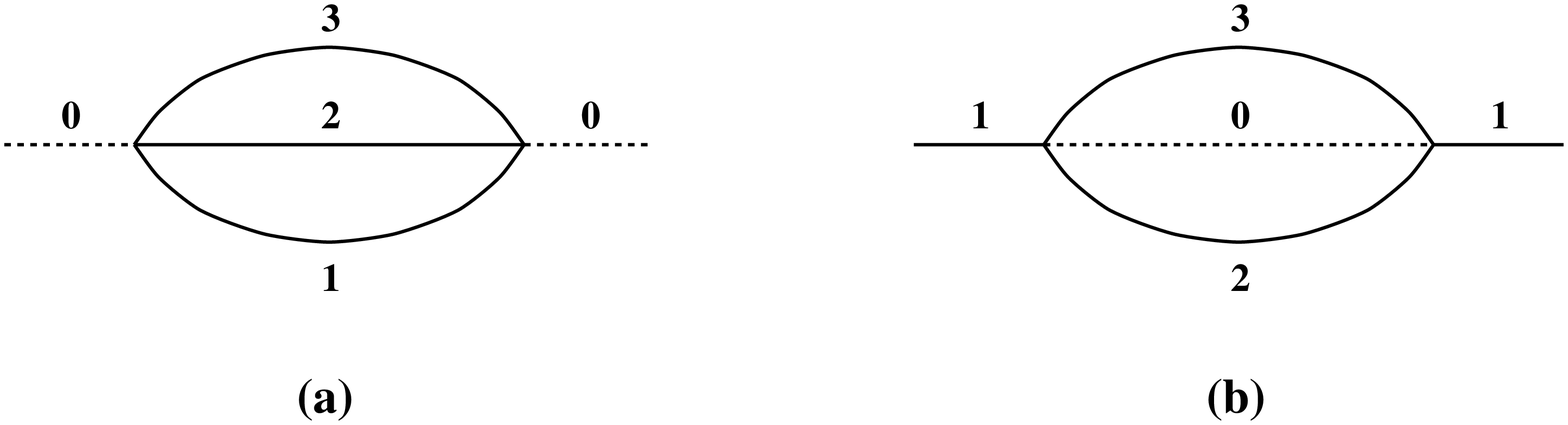}
\caption{
}
\label{fig:5}       
\end{figure}

\begin{figure}
\epsfxsize=5.0in
\hspace*{0.0cm}\epsffile{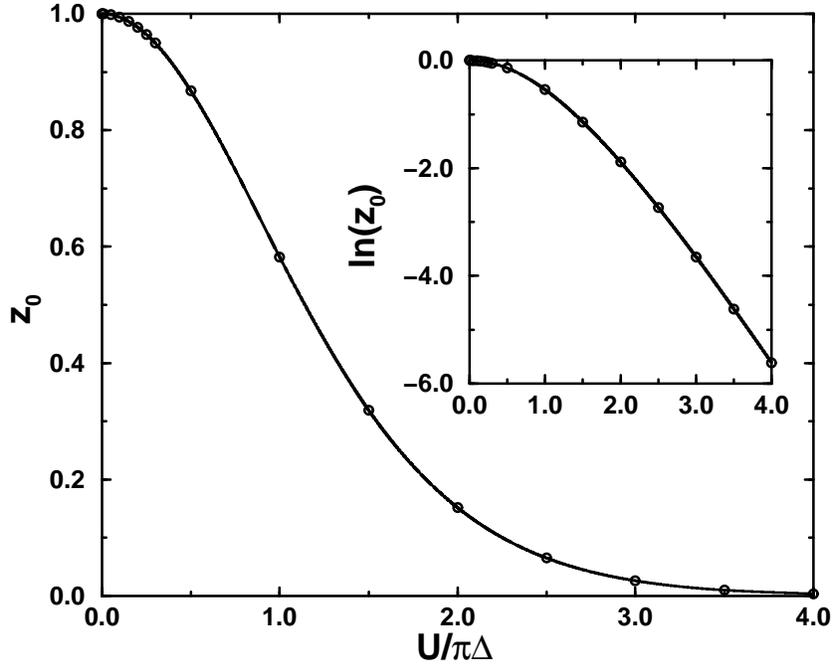}
\caption{The wavefunction renormalization factor, 
\( z_0 \), determined by the coefficient of \( 
\delta \left( \omega \right) \) in the spectral
 density, \( \rho_{0} \left( \omega \right) \).
 Exponential behaviour for large \( U/\pi\Delta 
\) is indicated by the linear region of the 
inset \( {\rm ln}  \left( z_0 \right) \) plot.}
\label{fig:6}       
\end{figure}

\begin{figure}
\epsfxsize=5.0in
\hspace*{0.0cm}\epsffile{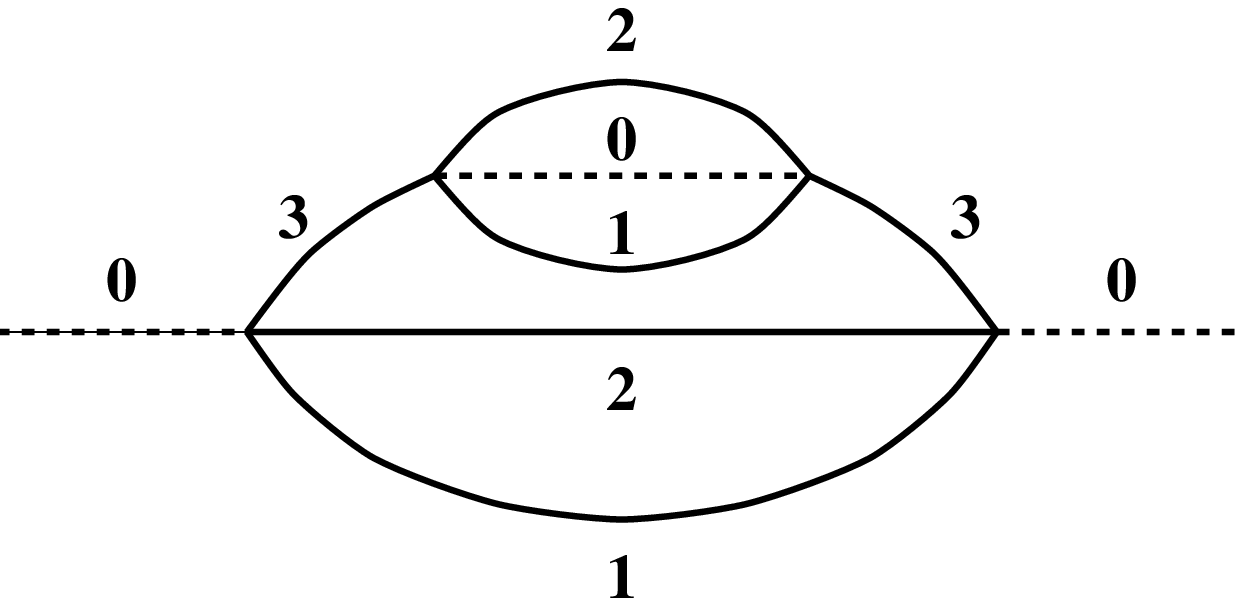}
\caption{
}
\label{fig:7}       
\end{figure}

\begin{figure}
\epsfxsize=4.5in
\hspace*{0.0cm}\epsffile{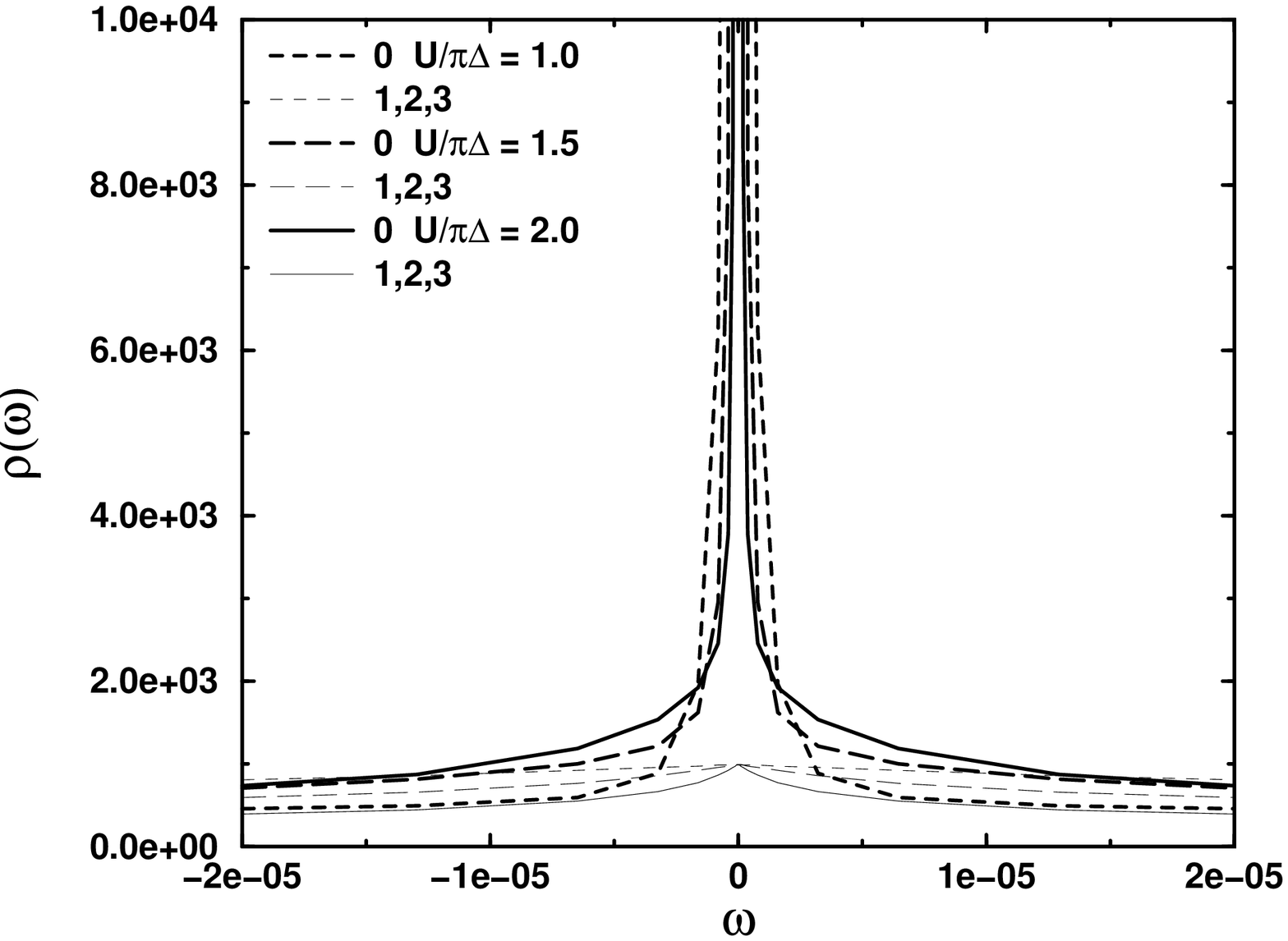}
\caption{The spectral density \( \rho \left( \omega 
\right) \) of the Majorana fermion impurity Green's 
function in the Fermi liquid situation: 
\( V=0.01414, V_0=0.00014 \ \left( \Delta 
= 10^{-4}\pi, \Delta_0 = 10^{-8}\pi \right) \).}
\label{fig:8}       
\end{figure}

\begin{figure}
\epsfxsize=4.5in
\hspace*{0.0cm}\epsffile{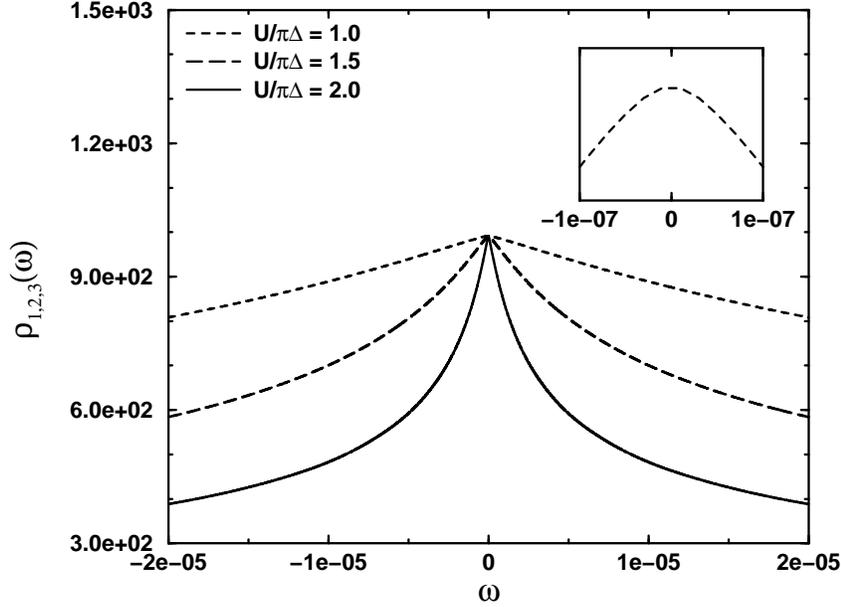}
\caption{The spectral density \( \rho_{1,2,3} 
\left( \omega \right) \) of the \( \alpha=1,2,3 \) 
Majorana fermion Green's function in the Fermi 
liquid situation: \( V=0.01414, V_0=0.00014 \ 
\left( \Delta = 10^{-4}\pi, \Delta_0 = 
10^{-8}\pi \right) \). The inset shows 
the region of the spectrum near the Fermi 
level where flattening consistent with \( \omega^2 \) 
behaviour is observed. \( \rho_{1,2,3} \left( 0 \right) 
\) is within 2\% of the value given by the Friedel 
sum rule.}
\label{fig:9}       
\end{figure}

\begin{figure}
\epsfxsize=5.0in
\hspace*{0.0cm}\epsffile{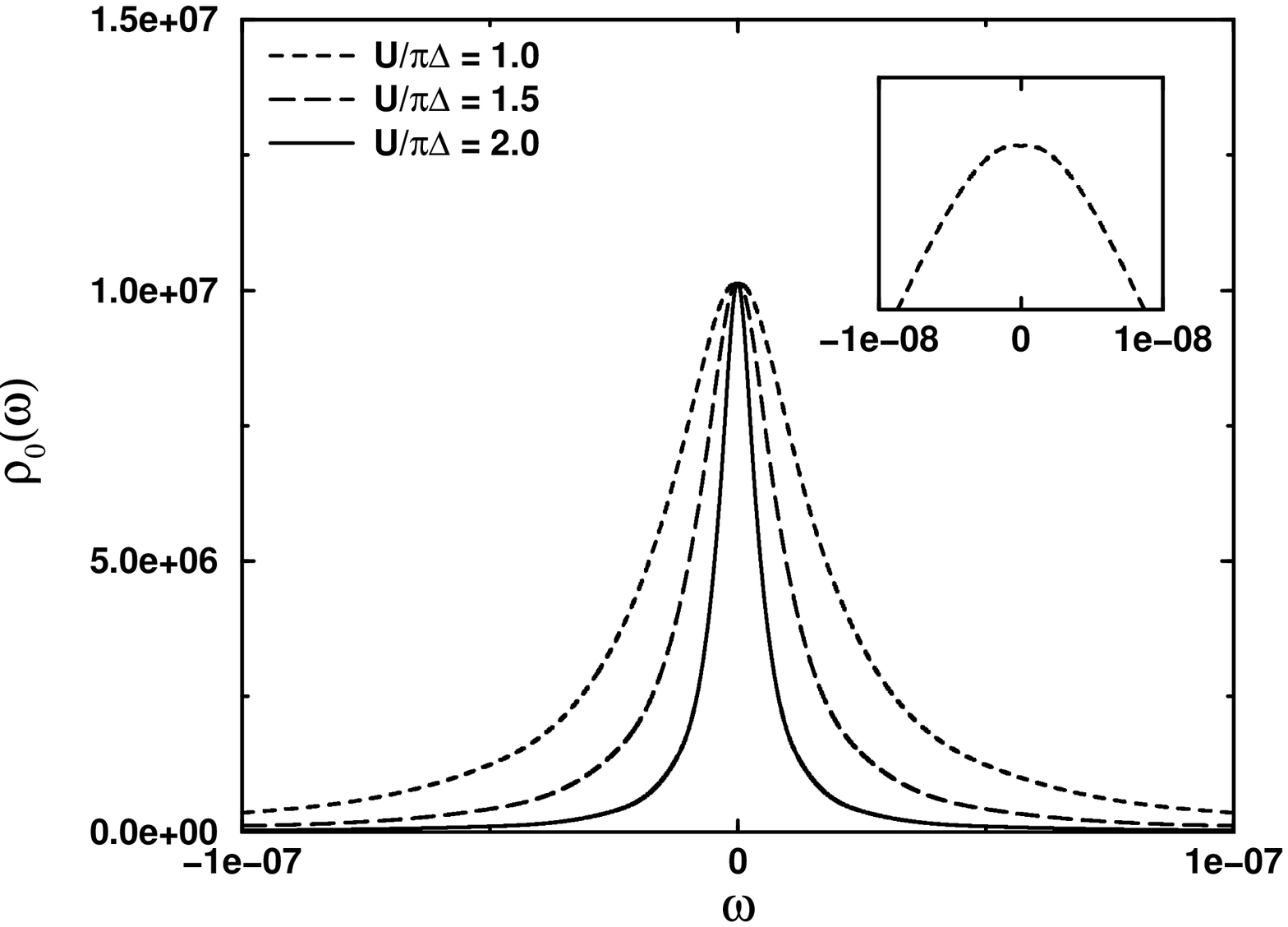}
\caption{The spectral density \( \rho_{0} \left( 
\omega \right) \) of the \( \alpha=0\) Majorana 
fermion Green's function in the Fermi liquid 
situation: \( V=0.01414, V_0=0.00014 \ \left( 
\Delta = 10^{-4}\pi, \Delta_0 = 10^{-8}\pi 
\right) \). The inset shows the \( \omega^2 \) 
dependence near the Fermi level. \( \rho_{0} 
\left( 0 \right) \) is within 2\% of the value 
given by the Friedel sum rule. }
\label{fig:10}       
\end{figure}

\begin{figure}
\epsfxsize=5.0in
\hspace*{0.0cm}\epsffile{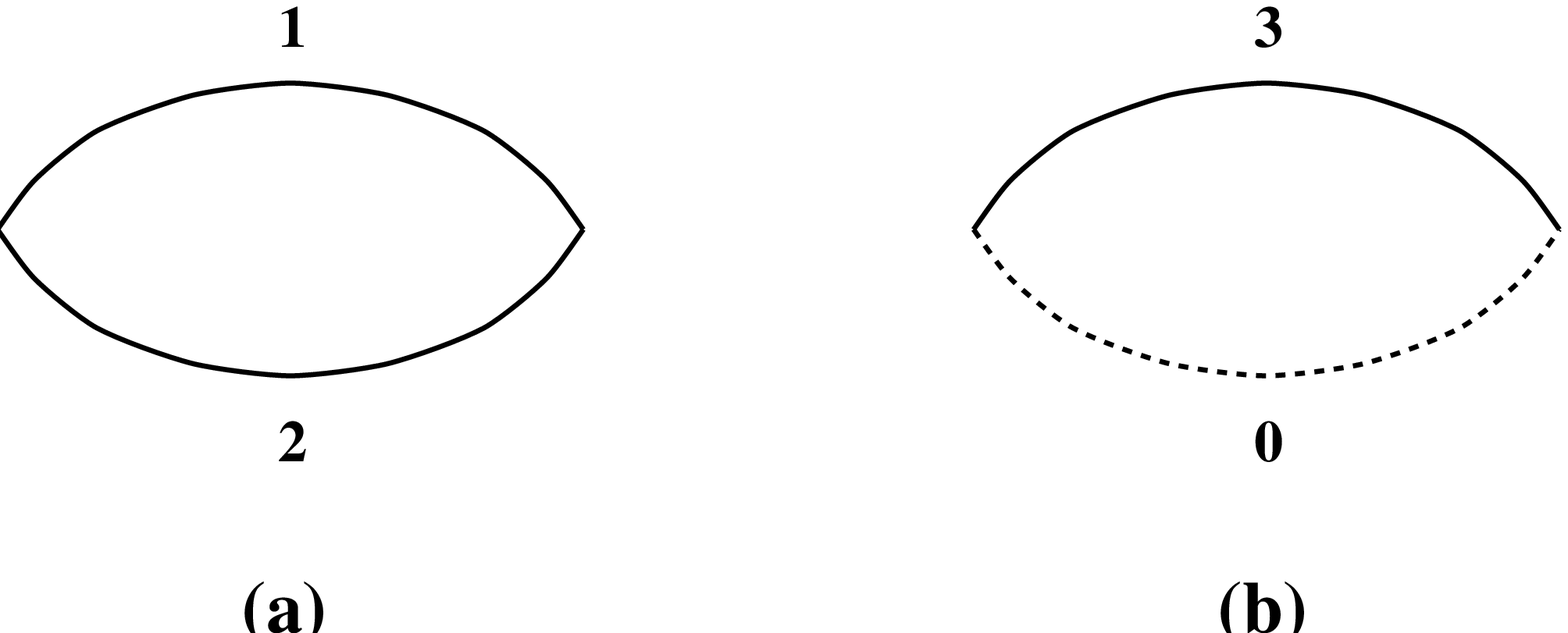}
\caption{
}
\label{fig:11}       
\end{figure}

\begin{figure}
\epsfxsize=5.0in
\hspace*{0.0cm}\epsffile{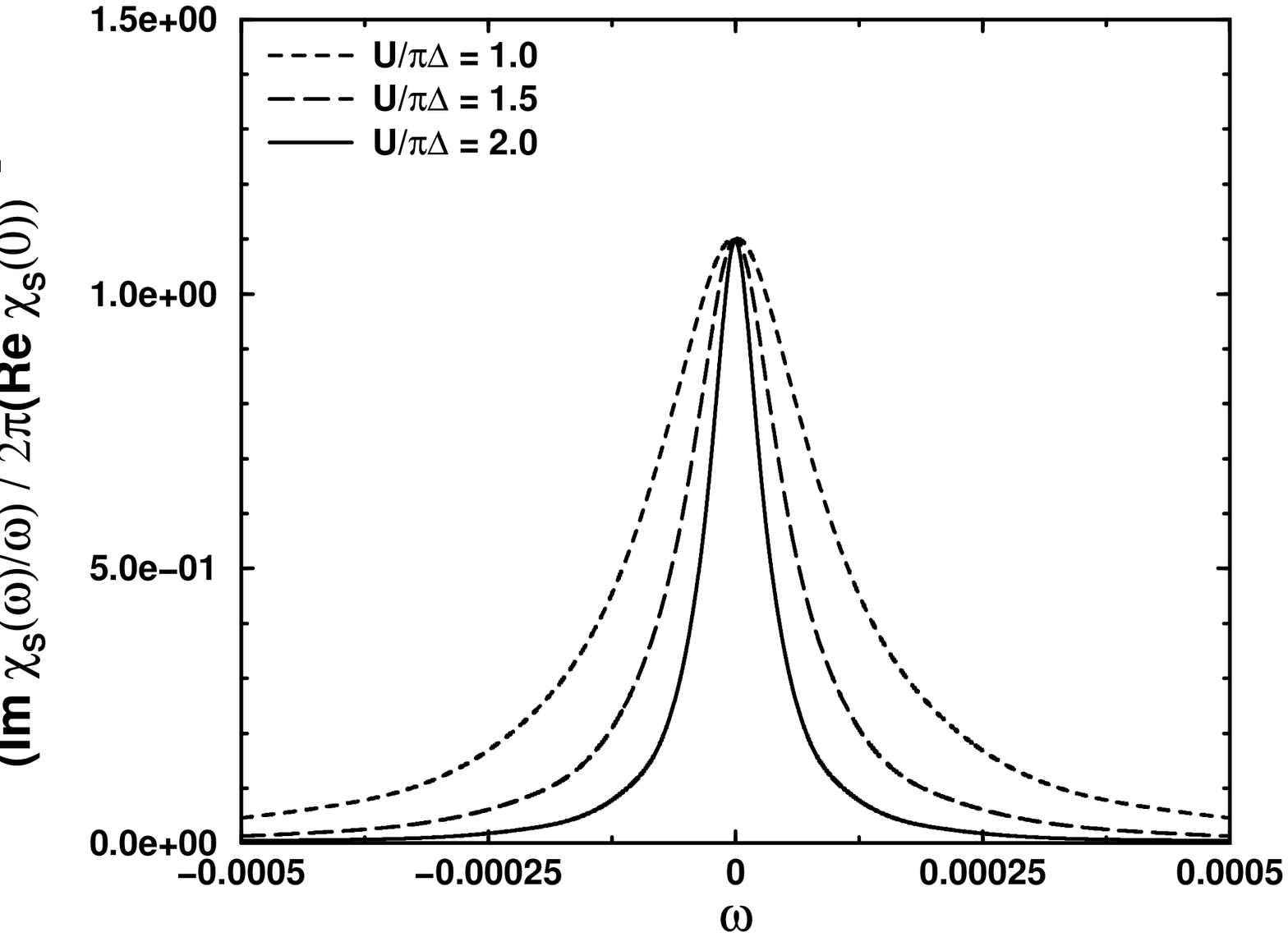}
\caption{\( \left( {\rm Im}\, \chi_s \left( \omega 
\right) / \omega  \right) / 2 \pi \left( {\rm Re}\, 
\chi_s \left( 0 \right) \right)^2 \) for the standard
 model: \( V=V_0=0.01414\ \) \(\left( \Delta = 
\Delta_0 = 10^{-4}\pi \right) \).}
\label{fig:12}       
\end{figure}

\begin{figure}
\epsfxsize=5.0in
\hspace*{0.0cm}\epsffile{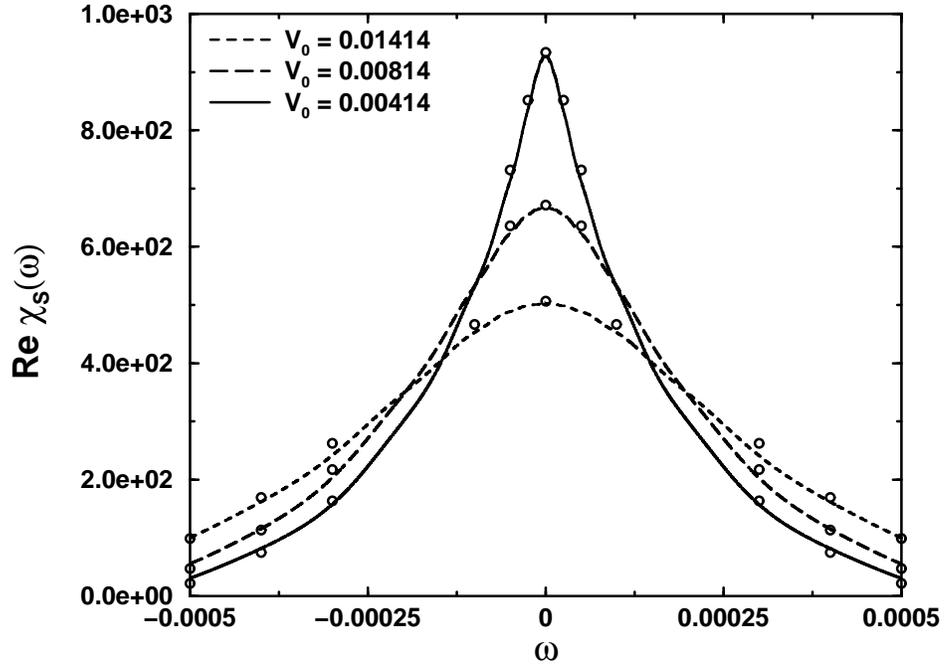}
\caption{The real part of the dynamic spin 
susceptibility \( \chi_s \left( \omega \right) 
\) for the non-interacting model: \( V=0.01414\ 
\) \( \left( \Delta = 10^{-4}\pi \right) \), 
\ \(U/\pi\Delta = 0.0\). The circles denote analytic 
values.}
\label{fig:13}       
\end{figure}

\begin{figure}
\epsfxsize=5.0in
\hspace*{0.0cm}\epsffile{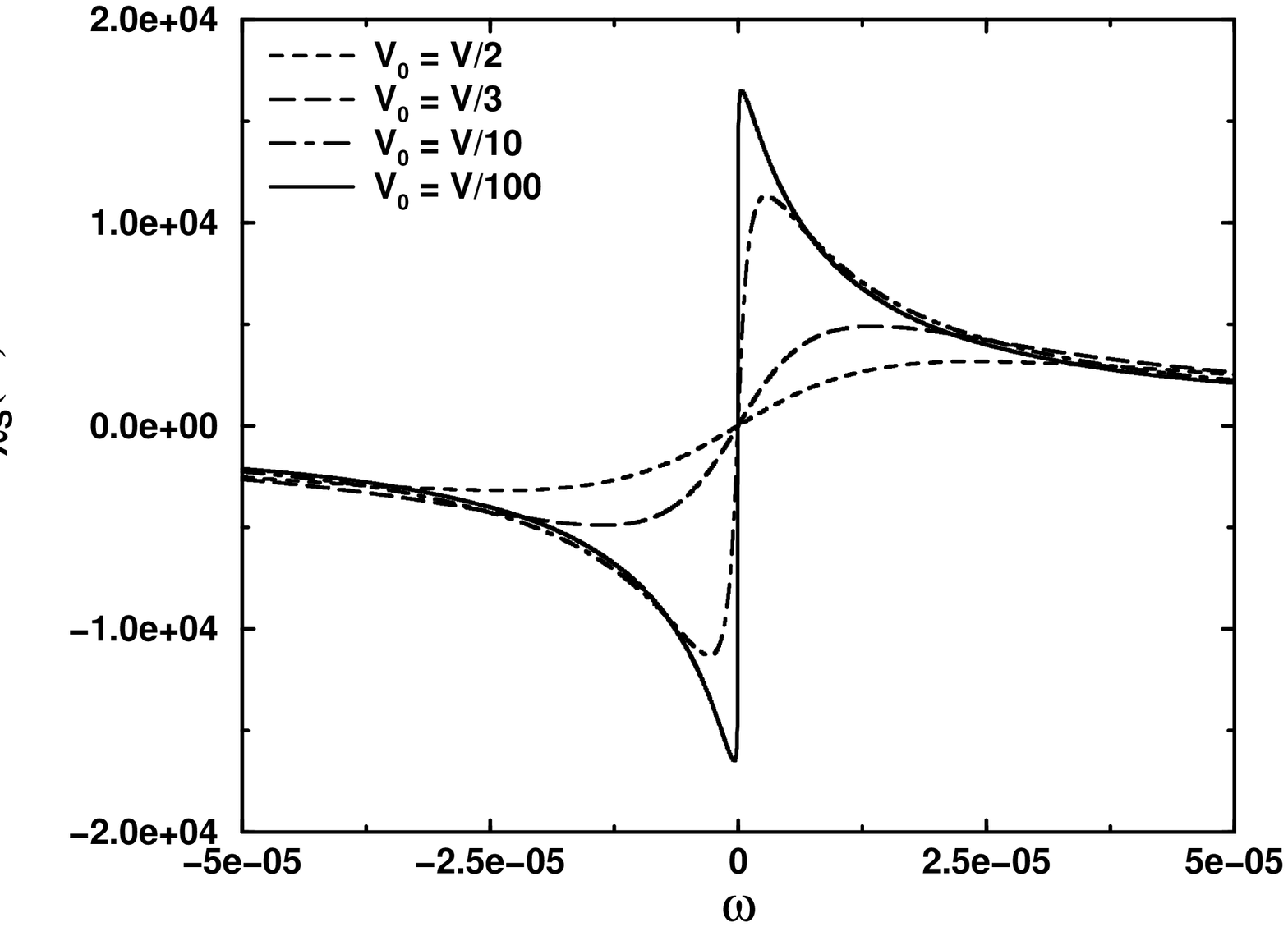}
\caption{The imaginary part of the dynamic spin 
susceptibility \( \chi_s \left( \omega \right) 
\) in the Fermi liquid situation: \( V=0.01414 
\ \left( \Delta = 10^{-4}\pi \right), \ 
U/\pi\Delta = 1.5\).}
\label{fig:14}       
\end{figure}

\begin{figure}
\epsfxsize=5.0in
\hspace*{0.0cm}\epsffile{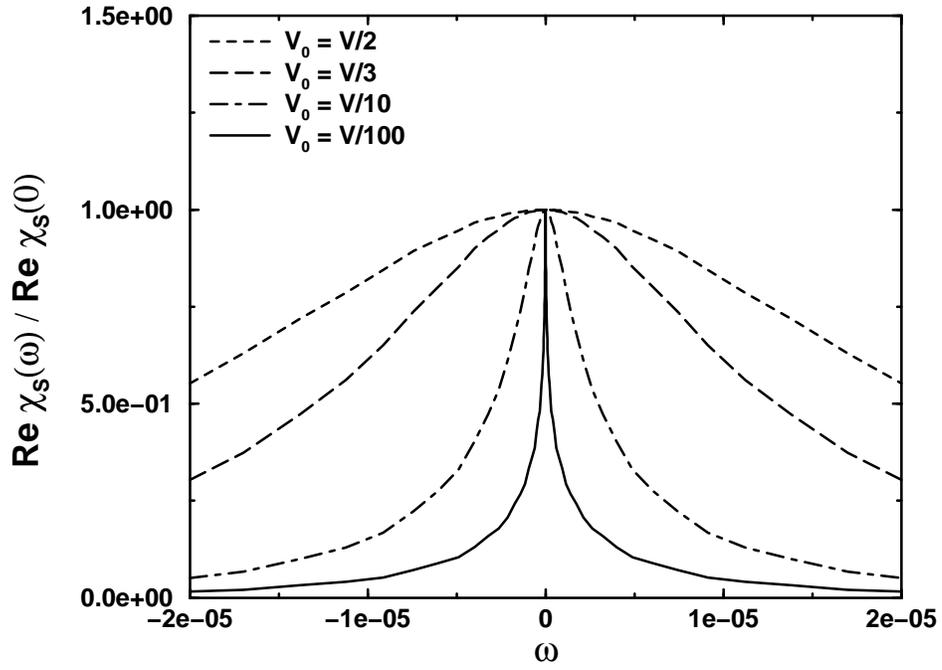}
\caption{The real part of the dynamic spin 
susceptibility \( \chi_s \left( \omega \right) 
/ \chi_s \left( 0 \right)\) in the Fermi 
liquid situation: \( V=0.01414 \ \left( 
\Delta = 10^{-4}\pi \right), \ U/\pi\Delta = 1.5\).}
\label{fig:15}       
\end{figure}

\begin{figure}
\epsfxsize=5.0in
\hspace*{0.0cm}\epsffile{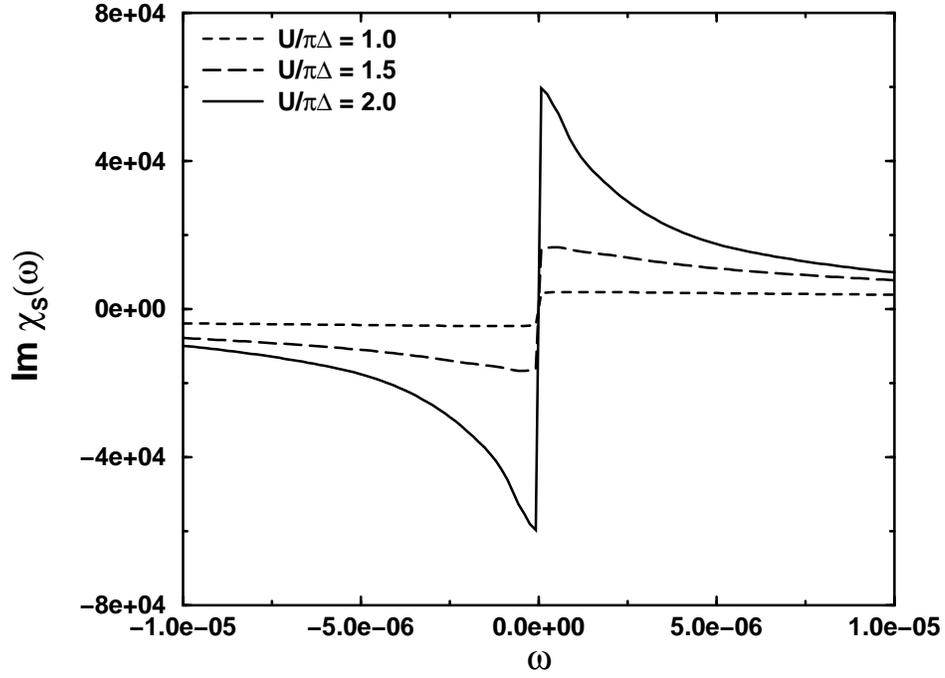}
\caption{The imaginary part of the dynamic spin 
susceptibility \( \chi_s \left( \omega \right) 
\) in the marginal Fermi liquid situation: 
\( V=0.01414, V_0=0.0 \ \left( \Delta = 
10^{-4}\pi, \Delta_0 = 0.0 \right) \).}
\label{fig:16}       
\end{figure}

\begin{figure}
\epsfxsize=5.0in
\hspace*{0.0cm}\epsffile{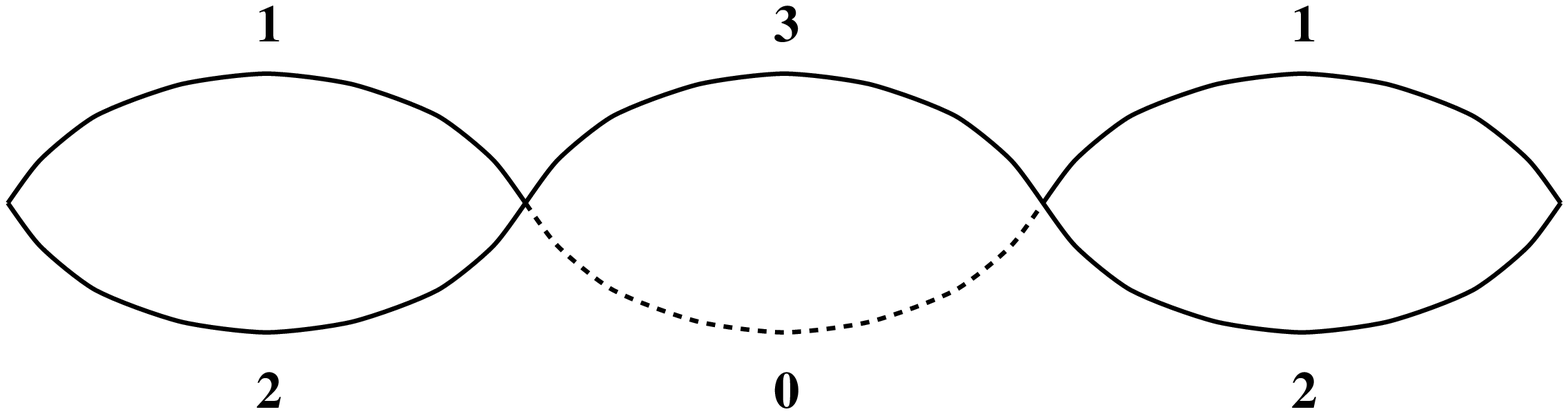}
\caption{
}
\label{fig:17}       
\end{figure}

\begin{figure}
\epsfxsize=5.0in
\hspace*{0.0cm}\epsffile{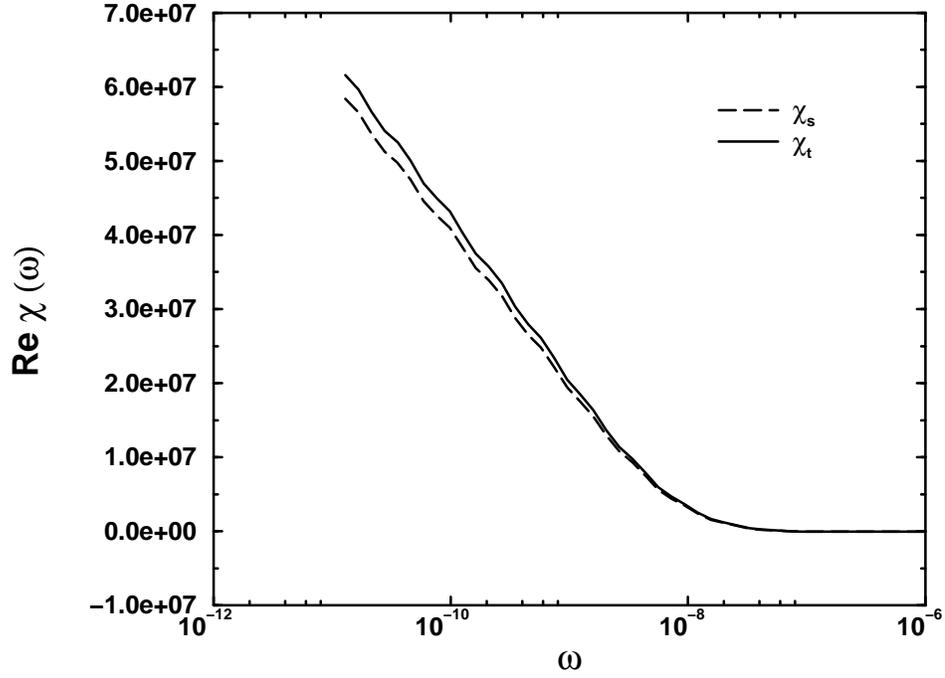}
\caption{The real part of the dynamic spin 
susceptibilities \( \chi_{\bf t} \left( 
\omega \right) \) and \( \chi_s \left( 
\omega \right) \)in the marginal Fermi 
liquid situation: \( V=0.01414, V_0=0.0 
\ \left( \Delta = 10^{-4}\pi, \Delta_0 = 
0.0 \right), \ U/\pi\Delta = 4.0 \), 
displaying linear dependence on \( 
{\rm ln}\left( \omega \right) \) 
in the low energy region.}
\label{fig:18}       
\end{figure}

\end{center}

\end{document}